# Recurrence Quantification Analysis of Dynamic Brain Networks

Marinho A. Lopes[1,2,*], Jiaxiang Zhang[2], Dominik Krzemiński[2], Khalid Hamandi[2], Qi Chen[3], Lorenzo Livi[4,5], and Naoki Masuda[6,7,1#]

[1]Department of Engineering Mathematics, University of Bristol, BS8 1UB, United Kingdom

[2]Cardiff University Brain Research Imaging Centre, School of Psychology, Cardiff University, Cardiff CF24 4HQ, United Kingdom

[3]Center for Studies of Psychological Application and School of Psychology, South China Normal University, Guangzhou 510631, China

[4]Departments of Computer Science and Mathematics, University of Manitoba, Winnipeg, MB R3T 2N2, Canada

[5]Department of Computer Science, College of Engineering, Mathematics and Physical Sciences, University of Exeter, Exeter EX4 4QF, United Kingdom

[6]Department of Mathematics, University at Buffalo, State University of New York, USA

[7]Computational and Data-Enabled Science and Engineering Program, University at Buffalo, State University of New York, USA

*m.lopes@exeter.ac.uk

#naokimas@buffalo.edu

**Abstract**

Evidence suggests that brain network dynamics are a key determinant of brain function and dysfunction. Here we propose a new framework to assess the dynamics of brain networks based on recurrence analysis. Our framework uses recurrence plots and recurrence quantification analysis to characterize dynamic networks. For resting-state magnetoencephalographic dynamic functional networks (dFNs), we have found that functional networks recur more quickly in people with epilepsy than healthy controls. This suggests that recurrence of dFNs may be used as a biomarker of epilepsy. For stereo electroencephalography data, we have found that dFNs involved in epileptic seizures emerge before seizure onset, and recurrence analysis allows us to detect seizures. We further observe distinct dFNs before and after seizures, which may inform neurostimulation strategies to prevent seizures. Our framework can also be used for understanding dFNs in healthy brain function and in other neurological disorders besides epilepsy.

# 1. Introduction

The brain is a complex dynamic system. To map, model and study brain structure and function, it is useful to define brain networks (Fornito et al., 2016; Bassett and Sporns, 2017). The study of brain networks has transformed our understanding of the brain, and it has the potential to revolutionize the clinical management of neurological disorders (Stam, 2014). Two main types of brain networks have been considered: structural and functional networks (Bullmore and Sporns, 2009; Sporns 2013). Structural networks describe the anatomical connectivity of the brain and are relatively stable on short time scales (i.e., seconds to minutes). In contrast, functional networks are inferred from statistical dependencies between neural signals recorded from different brain regions. Statistical dependencies are then assumed to represent functional couplings between brain regions. The statistical dependencies between signals are not stationary, making functional networks time-dependent on short time scales (tens or hundreds of milliseconds) (Sporns, 2013). A growing body of evidence shows that dynamic functional networks (dFNs) capture crucial aspects underlying normal function and dysfunction of the brain (Hutchison et al., 2013; Calhoun et al., 2014; Cohen, 2018). In particular, epilepsy, which will be the focus of the present study, has been considered to be a dynamical disease of the brain (da Silva, 2003), and dFNs have been useful for characterizing the epileptic brain (Lehnertz et al., 2014).

A number of different approaches have been employed to study dFNs (Braun et al., 2018; Hutchison et al., 2013). A common approach has been to calculate some measures from the functional networks and track their changes over time (Schindler et al., 2008; Kramer et al., 2010; Lehnertz et al., 2014; Fuertinger et al., 2016). For example, a time-dependent analysis of the average shortest path length and the clustering coefficient revealed that functional networks evolve from a more random topology before seizures towards a more regular topology during seizures and back to a more random topology after seizure offset (Schindler et al., 2008). This approach is limited by an a priori choice of measures that may or may not fully characterize the dynamics of the functional networks. Another common approach is to use a Bayesian framework to characterize dFNs. In particular, hidden Markov models have been employed to analyze dFNs (Eavani et al., 2013; Sourty et al., 2016; Vidaurre et al., 2018). For example, product hidden Markov models have been used to identify brain networks involved in dementia with Lewy bodies (Sourty et al., 2016). This approach makes the assumption that the dynamics of the state is Markovian, i.e. the transition between different states is a memoryless stochastic process. However, long-term correlations in temporal patterns of brain activity suggest that this assumption may not always hold (Kitzbichler et al., 2009; Chialvo, 2010; Ezaki et al., 2019). Thus, these approaches make assumptions that may hinder a comprehensive assessment of the dynamics of functional networks.

Recurrence is a key concept in dynamical systems (Eckmann et al., 1995; Marwan et al., 2007). It was first introduced by Henri Poincaré in 1890 and it can be used to characterize a system's dynamical behavior. In the late 1980s, Eckmann et al. introduced the *recurrence plot* (RP), a tool to visualize the recurrences of a dynamical system (Eckmann et al., 1987). Subsequently, *recurrence quantification* *analysis* (RQA) emerged as a means to quantify RPs (Marwan et al., 2007). RQA has been applied to a range of dynamical systems and empirical data (Marwan et al., 2007; Webber and Marwan, 2015). In particular, RQA has been used for examining brain activity (see e.g. Ouyang et al., 2008; Shabani et al., 2016; Ngamga et al., 2016; Yan et al., 2016; Yang et al., 2019). RQA has been used for the identification of pre-seizure states from intracranial EEG data recorded from people with epilepsy (Ngamga et al., 2016), and for distinguishing EEG signals between healthy individuals and people with epilepsy (Yan et al., 2016). More recently, RQA indicated that the epileptogenic zone produces more deterministic dynamics than other brain regions (Yang et al., 2019). However, these approaches, including the application of RP and RQA to other neural data, neglect the spatial dependencies between the neural signals, i.e. they do not assess the recurrence of the underlying functional networks.

In the present study, we propose methods of RP and RQA for dFNs. While the methods may be used to explore dFNs from various neurological disorders, here we focus on epilepsy. Functional networks may be crucial for understanding epilepsy (Richardson, 2012). Since functional brain networks are dynamic and epilepsy is a dynamical disease, RQA of dynamical functional networks may be particularly useful to characterize the epileptic brain. We use two data sets. The first one comprises MEG resting-state signals from people with a generalized epilepsy syndrome, Juvenile Myoclonic Epilepsy (JME) and healthy controls. Our aim is to test whether dFNs differ between the two groups, and whether this can be used as a biomarker of JME. The second data set comprises invasive stereo EEG (sEEG) recordings from people with drug-resistant focal epilepsy. Using this data set, we aim to assess functional network dynamics before, during and after seizures, particularly examining whether functional networks before seizures recur during and after seizures and across different seizures. We also assess whether RQA of dFNs may be used to automatically detect seizures. The application of our methods to these two data sets allows us to show the methods' flexibility and versality, which in turn enables us to test the range of different hypotheses just described.

## 2. Materials and Methods

Our framework to study dynamic brain networks comprises five key steps (Fig. 1): (i) use a sliding window approach to segment data from MEG or EEG recordings (Fig. 1(a)); (ii) infer a functional network from each segment of data (Fig. 1(b)); (iii) compute the distance between pairs of functional networks (Fig. 1(c)); (iv) apply a threshold to the pairwise distances and obtain a recurrence plot (RP) (Fig. 1(d)); and (v) perform recurrence quantification analysis (RQA) to extract information from the RP (Fig. 1(e)).

### 2.1. Data acquisition and pre-processing

We study two data sets from patients with epilepsy: one comprising MEG recordings and the other containing stereo EEG (sEEG) recordings.

### 2.1.1. MEG recordings

We consider resting-state MEG data recorded from 26 people with JME and 26 healthy controls. The people with epilepsy were recruited from a specialist clinic for epilepsy at University Hospital of Wales in Cardiff, and the healthy controls were volunteers who had no history of significant neurological or psychiatric disorders. The two groups were age and gender matched (age range [19,45], median 27 years, and 8 males in the epilepsy group; age range [18,48], median 27, and 7 males in the control group). People with epilepsy had a number of different seizure types and were taking anti-epileptic drugs (see Krzemiński et al., 2019 for more details about this dataset). MEG data were acquired using a 275-channel CTF radial gradiometer system (CTF System, Canada) at a sampling rate of 600 Hz. Recording sessions lasted for approximately 5 minutes per individual. The participants were instructed to sit steadily in the MEG chair with their eyes focused on a red dot on a grey background. The participants also underwent a whole-brain T1-weighted MRI acquired using a General Electric HDx 3T MRI scanner and an 8-channel receiver head coil (GE Healthcare, Waukesha, WI) with an axial 3D fast spoiled gradient recalled sequence (echo time 3 ms; repetition time 8 ms; inversion time 450 ms; flip angle 20º; acquisition matrix 256×192×172; voxel size 1×1×1 mm). This study was approved by the South East Wales NHS ethics committee, Cardiff and Vale Research and Development committees, and Cardiff University School of Psychology Research Ethics Committee. Written informed consent was obtained from all participants.

The data was first divided into 2 s segments and then each segment was visually inspected to remove motion, muscle and eye-blink artefacts, and also interictal spike wave discharges from the MEG recordings. Artefact-free segments were identified and re-concatenated for each subject. The resulting epochs had variable lengths ranging from 204 s to 300 s. The pre-processed data were then filtered in the classical frequency bands (theta (4-7 Hz), alpha (8-13 Hz), beta (15-25 Hz) and gamma (30-60 Hz) bands), and down-sampled to 250 Hz.

### 2.1.2. sEEG recordings

We also used a data set comprising 10 people with drug-resistant focal epilepsy who underwent invasive monitoring with stereo EEG at the 999 Brain Hospital, China. Stereo EEG is an advanced procedure in the epilepsy surgery evaluation, to help delineate the irritative and seizure onset zones, and hence decide the suitability and plan epilepsy surgery (Duncan et al., 2016). The age range of the group was [16,31], median 23, and 9 individuals were males. Electrode implantation locations were personalized according to imaging and non-invasive EEG data; the number of electrodes per implantation ranged from 5 to 16, and each electrode had 2 to 16 contacts (i.e., channels). Four individuals received bilateral implantations, two individuals had electrodes implanted in their right hemispheres, and the other four individuals received implantations in their left hemispheres. Stereo EEG data was acquired using Nihon Kohden recording system at a sampling rate of 1 or 2 kHz. For all individuals, high-resolution T1-weighted MRI were acquired before electrode implantation, and computed tomography (CT) were acquired after electrode implantation. Co-registration of the CT to the MR images allowed us to determine whether contacts were placed in grey matter. All individuals had at least 2 seizures recorded. Seizure onset was defined by epileptologists at the 999 Brain Hospital and corroborated by one of the authors (KH) who also marked seizure offset. This study was approved by 999 Brain Hospital ethics committee and South China Normal University ethics committee. Written informed consent was obtained from all participants.

We restricted our analysis to artefact-free sEEG channels placed on grey matter as established from co-registration of the CT scans to the MRIs. (Artefact-free channels were identified by KH.) The number of channels $N_{\text{ch}}$ considered per individual ranged from 19 to 83 (median 63.5). We selected peri-ictal epochs of data containing 300 s before seizure onset (pre-ictal), seizure (ictal), and 300 s after seizure offset (post-ictal). We then neglected peri-ictal epochs whose pre-ictal 300 s overlapped with the post-ictal 300 s of the previous peri-ictal epoch. Thus, we end up with 2 to 4 peri-ictal epochs per individual (3 individuals had 2 epochs each; 5 individuals had 3 epochs each; and 2 individuals had 4 epochs each, making a total of 29 peri-ictal epochs). The data was re-referenced to the average of all artefact-free channels, filtered in a broad frequency band (0.5-120 Hz), which encapsulates the traditional clinical frequency bands (delta, theta, alpha, beta, and gamma (Buzsaki, 2006)), notch filtered to remove power line interference (48 to 52 Hz), and down-sampled to 250 Hz.

### 2.2. Inferring functional networks

The first step of the method is to construct dFNs. In this section we describe how we segmented the pre-processed MEG and sEEG data and inferred a functional network from each segment.

### 2.2.1. Dynamic MEG functional networks in the source space

To compute functional networks from the MEG data, we first transformed the MEG recordings from the sensor space to the source space. This procedure consisted of co-registering the MEG sensors with the structural MRI using the locations of the fiducial coils in the CTF software (MRIViewer and MRIConverter). Then, we inferred a volume conduction model from the MRI scan using a semi-realistic model (Nolte, 2003). Finally, we reconstructed the source signals from the sensor signals using a linear constrained minimum variance (LCMV) beamformer on a 6-mm template with a local-spheres forward model in Fieldtrip (Oostenveld et al., 2011; http://www.ru.nl/neuroimaging/fieldtrip). Sources were mapped into the 90 brain regions of the Automated Anatomical Label (AAL) atlas (Hipp et al., 2012). More details about these methods were provided in our previous study (Krzemiński et al., 2019).

To compute dFNs, we divided the 90-dimensional source reconstructed MEG signals into segments, each of which was composed of 500 samples (i.e., 2 s). Each segment was subsequently used for constructing one functional network. The choice of segment length balances the need of a sufficient number of samples to infer a reliable functional network and the need of a sufficiently large number of functional networks, $M$, for analyzing their recurrence over time. The segment size of 500 is within a typical range in both MEG and EEG studies of functional connectivity (see e.g. Khambhati et al., 2015; Colclough et al., 2016; Stahn and Lehnertz, 2017). We set the overlap between consecutive segments to be 80% such that consecutive segments shared 400 samples, i.e., 1600 ms. Therefore, the time step between consecutive dFNs was 100 samples, i.e. 400 ms. This choice represents a compromise between the need of sufficiently many networks for subsequent analysis, which is satisfied with a large overlap, and the need of avoiding trivial recurrences between consecutive functional networks. See Supplementary Material S1 for computational results underlying the choice of the 80% overlap. The MEG data had different lengths for different participants. Therefore, we only considered the first $M = 506$ segments, which was the minimum number of segments among all participants, to exclude the potential impact of different recording lengths on our results.

For each segment, we built two functional networks using two established methods (Colclough et al., 2016): phase lag index (PLI) (Stam et al., 2007) and amplitude envelope correlation (AEC) with orthogonalized signals (Hipp et al., 2012) (see Supplementary Material S2 for more details). We selected the PLI and AEC because they measure different types of intrinsic coupling, one related to phase coupling, and the other to amplitude correlations. They are expected to complement each other in describing brain network interactions (Engel et al., 2013, Guggisberg et al., 2015). Note that since we considered 4 frequency bands for each definition of functional connectivity (i.e., PLI and AEC), we obtained eight sequences of matrices $A(t) = (a_{ij}(t))$, where $i, j = 1, \ldots, 90$, and $t = 1, \ldots, 506$. Each matrix $A(t)$ represented a functional network for segment $t$, and matrix entry $a_{ij}(t)$ represented the strength of the functional connectivity between regions $i$ and $j$. Each matrix $\boldsymbol{A}(t)$ was symmetric, i.e., the functional networks were undirected.

*2.2.2. Dynamic sEEG functional networks in the sensor space*

Following the same procedure as the one employed for the MEG data, we divided each of the 29 sEEG peri-ictal epochs into segments of 500 samples (i.e., 2 s) using an overlap of 400 samples between consecutive segments. Because different peri-ictal epochs contained seizures of different lengths (from 9 to 181 s, median 75 s), different peri-ictal epochs resulted in different numbers of segments $M$ (from 1542 to 1822, median 1632).

Because the sEEG data was recorded intracranially close to the brain sources, we computed functional networks in the sensor space, i.e. using the channels as network nodes and functional connections as the statistical dependencies between the signals recorded at the channels. For each segment, we inferred a functional network using the absolute value of the Pearson's correlation coefficient between pairs of channels (Rummel et al., 2010; Rummel et al., 2015; Lopes et al., 2017). These methods differ from those applied to the MEG data because the two data modalities are different with regard to volume conduction. Volume conduction is a confounding factor in non-invasive data and is responsible for spurious zero-lag correlations (Bastos and Schoffelen, 2016). Thus, while we had to use methods capable of accounting for volume conduction in the MEG data set, such a concern does not apply to invasive sEEG data, for which the Pearson's correlation based on broadband signals is a reliable method (Rummel et al., 2010). The Pearson's correlation is considered to be the simplest measure to capture possible linear relationships between two signals (Bastos and Schoffelen, 2016). Thus, we obtained 29 time-varying matrices $A(t) = (a_{ij}(t))$ of size $(N_{\text{ch}} \times N_{\text{ch}})$, where $t = 1, \ldots, M$, representing functional networks through pre-ictal, ictal, and post-ictal periods. Note that the number of channels, $N_{\text{ch}}$, was fixed for each individual, but $M$ varied from one peri-ictal epoch to another even for a single individual due to seizure of different lengths.

## 2.3. Recurrence plots and distance measures

The second step of the method is to obtain an RP from a dFN.

For a dynamical system characterized by a vector time series $\vec{x}(t)$, where $t = 1, \ldots, M$, an $M \times M$ recurrence matrix is defined as

$$R_{t_1, t_2} = \begin{cases} 1 & \text{if } d(\vec{x}(t_1), \vec{x}(t_2)) \leq \epsilon, \\ 0 & \text{if } d(\vec{x}(t_1), \vec{x}(t_2)) > \epsilon, \end{cases} \quad (1)$$

where $d(\vec{x}(t_1), \vec{x}(t_2))$ is a distance measure between $\vec{x}(t_1)$ and $\vec{x}(t_2)$, and $\epsilon$ is a small distance which defines an upper limit of discrepancy between recurrent states (Marwan et al., 2007; Marwan et al., 2009). The recurrence matrix is a symmetric matrix, and its main diagonal entries are equal to 1.

To compute RPs of dFNs, we replaced the vectors $\vec{x}(t_a)$ by matrices $A(t_a)$ and used distance measures for pairs of weighted networks (i.e., matrices). The recurrence matrix for a dFN is given by

$$R_{t_1,t_2} = \begin{cases} 1 & \text{if } d(A(t_1), A(t_2)) \leq \epsilon, \\ 0 & \text{if } d(A(t_1), A(t_2)) > \epsilon, \end{cases} \quad (2)$$

where $d(A(t_1), A(t_2))$ is the distance between functional networks $A(t_1)$ and $A(t_2)$ measured according to a distance measure for networks. There is a variety of distance measures for networks (Livi and Rizzi, 2013), but a good choice for functional networks is unknown. We therefore used six distance measures to obtain six different RPs per dFN. The use of multiple distance measures aimed at not missing potentially useful information provided by different, yet arbitrary choices of distance measures. We considered the Frobenius norm, the log-Euclidean distance, the spectral norm, the Euclidean norm between Fiedler vectors, the maximum norm between the Fielder vectors, and the cosine dissimilarity between the Fiedler vectors. We then assessed whether these distance measures were actually complementing each other or being redundant (we will return to this issue in section 2.5 below). We reduced the number of distance measures to three of interest: the Frobenius norm, the spectral norm, and the Euclidean norm between Fiedler vectors. These measures were applied to pairs of networks and are defined as follows.

The Frobenius norm of the difference between a pair of networks is given by (Kurmukov et al., 2016)

$$d_F\big(A(t_1), A(t_2)\big) = \|A(t_1) - A(t_2)\|_F = \sqrt{\sum_{i=1}^{N}\sum_{j=1}^{N}\big(a_{ij}(t_1) - a_{ij}(t_2)\big)^2}. \qquad (3)$$

This distance measure is the Euclidean distance between the two networks when they are represented as $M^2$-dimensional vectors.

The spectral norm of the difference between a pair of networks is given by

$$d_S\big(A(t_1), A(t_2)\big) = \sqrt{\lambda_{\max}\{[A^*(t_1) - A^*(t_2)][A(t_1) - A(t_2)]\}}, \qquad (4)$$

where $\lambda_{\max}\{\cdot\}$ is the largest eigenvalue of the matrix, and $A^*(t)$ is the conjugate transpose of $A(t)$ (Miller et al., 2012). In fact, our matrices $A(t)$ are real, and therefore $A^*(t)$ is just the transpose of $A(t)$.

The third measure is the Euclidean norm between Fiedler vectors, which, as the name suggests, is based on spectral properties of the Laplacian networks, specifically their Fiedler vectors. The Fiedler vector of a network is the eigenvector corresponding to the smallest positive eigenvalue of the Laplacian matrix of the network, which is often referred to as the algebraic connectivity of the network. The use of the Laplacian matrix is motivated by evidence showing that the Laplacian matrix is better for computing spectral distances than the adjacency matrix (Wilson and Zhu, 2008). The Fiedler vector characterizes the partitioning of the network into communities (Newman, 2006). Here, we considered the so-called symmetric normalized Laplacian matrix given by $L'(t_a) = D^{-1/2}(t_a)L(t_a)D^{-1/2}(t_a)$, where $L(t_a) = D(t_a) - A(t_a)$ is the combinatorial Laplacian, and $D(t_a)$ is a diagonal matrix whose main diagonal entries are given by $d_{ii}(t_a) = \sum_{j=1}^{N} a_{ij}(t_a)$. We denote the normalized Fiedler vector of $L'(t)$ by $\vec{v}(t) = (v_1(t), v_2(t), \ldots, v_N(t))^\top$, where ⊤ represents the transposition. To compute the distance between the normalized Fiedler vectors $\vec{v}(t_1)$ and $\vec{v}(t_2)$, we used the Euclidean norm between Fiedler vectors given by

$$d_{EF}(A(t_1), A(t_2)) = \sqrt{\sum_{k=1}^{N} (v_k(t_1) - v_k(t_2))^2}. \quad (5)$$

We used the appropriate orientation of the Fiedler vectors to calculate these distance measures (see the Supplementary Material S3 for more details). For details about the other three distance measures, see Supplementary Material S4.

To obtain an RP for each distance measure, one needs to define a threshold $\epsilon$. The value of $\epsilon$ may have a crucial impact on the structure of the RP (Marwan et al., 2007). We used $\epsilon$ such that the density of points in the RPs was fixed. In other words, all RPs had the same ratio of the number of recurrences to $M(M-1)$, the total number of possible recurrences. (Recurrence points along the main diagonal are ignored because they are trivial.) The advantage of this choice is the opportunity to compare the structure of different RPs, because such comparisons are only meaningful if the RPs have the same density of points (Marwan et al., 2007). We ran our analysis for three different thresholds such that the density of points was 0.01, 0.05, and 0.10, respectively.

## 2.4. Recurrence quantification analysis

The third step of the method is to quantify the structure of the RPs, which allows us to characterize the dynamics of the functional networks and compare different dFNs. For this purpose, we employed 12 RQA measures (Marwan et al., 2007), i.e., 11 out of the 13 measures implemented in version 5.22 of the CRP toolbox for MATLAB provided by TOCSY: http://tocsy.agnld.uni-potsdam.de, and the $\tau$-recurrence rate (denoted by $RR_\tau$). For a full description of the measures in the CRP toolbox, see Supplementary Material S6. We excluded the recurrence rate, i.e. the density of recurrence points in a RP, because we fixed this quantity to set the threshold, $\epsilon$, to build the RPs. We also excluded the clustering ($clust$) because it was generally small or undefined in our RPs due to the relatively low density of points in the RPs.

Among the 11 RQA measures in the CRP toolbox, four are based on the diagonal lines of the RPs, which result from recurring sequences of adjacent functional networks: the determinism ($DET$), the mean diagonal line length ($\langle L \rangle$), the maximal diagonal line length ($L_{\max}$), and the entropy of the diagonal line lengths ($ENTR$). In the analysis of the sEEG data set, we will highlight the $DET$, $\langle L \rangle$, and $L_{\max}$. A larger value of these measures implies that the dFNs are more "deterministic". Here, higher "determinism" means that the dFNs have longer consecutive sequences of functional networks that repeat at different times (Marwan et al., 2007).

Three other RQA measures quantify the vertical lines composed of recurrent points in the RPs, which represent time intervals in which the networks do not considerably change: the laminarity ($LAM$), the trapping time ($TT$), and the maximal vertical line length ($V_{\max}$). Among these measures, we will particularly focus on the $TT$ in section 3.1. It is equal to the average length of vertical lines in the RP. A large $TT$ value implies that the dFNs are more likely to be trapped into specific functional networks at any given time.

Another three RQA measures assess the recurrence times, i.e., the vertical distance between recurrence points in the RPs: the recurrence time of first type ($T1$), the recurrence time of second type ($T2$), and the recurrence time entropy ($RTE$). We highlight $T1$ and $T2$ in section 3.1. They quantify the average time that functional networks take to approximately recur to a previous network. The difference between $T1$ and $T2$ is that $T2$ neglects recurrence times equal to 1 which may correspond to spurious recurrences (see Supplementary Material S6 for more details).

The final RQA measure assesses the RP by regarding it as a network: the transitivity ($Trans$). These RQA measures quantify different aspects of the temporal dynamics enclosed in the RPs. We used the CRP toolbox provided by TOCSY to compute these RQA measures. These 11 measures were used to assess the MEG data set and to compare pre- and post-ictal periods in the sEEG data set.

The $RR_\tau$ was used to assess peri-ictal epochs in the sEEG data set. It is given by

$$RR_\tau = \frac{1}{M-\tau} \sum_{t=1}^{M-\tau} R_{t,t+\tau}. \tag{6}$$

This measure counts the number of recurrence points on diagonal lines with a distance $\tau$ from the main diagonal. The $RR_\tau$ can be considered as a generalized auto-correlation function (Marwan et al., 2007). To facilitate the interpretation of this measure, for each RP, we also computed 100 randomly shuffled RPs. We generated randomly shuffled RPs by taking the $(M-1)(M-2)/2$ matrix entries from the upper triangular part of the original RP matrix, uniformly randomly shuffling these entries, and then constructing a symmetric matrix using the shuffled upper triangular part and the main diagonal entries set to zero. Then, we computed the $RR_\tau$ value for each randomly shuffled RP, which we denote by $RR_\tau^{\text{null}}$. By calculating the mean and standard deviation of $RR_\tau^{\text{null}}$ at each $\tau$, based on the 100 samples, we obtained a reference to assess whether $RR_\tau$ deviated from chance at each $\tau$.

To assess the variation in RQA measures between pre- and post-ictal periods in the same peri-ictal epoch in the sEEG data, we computed $\Delta X = (X_{\text{post}} - X_{\text{pre}})/(X_{\text{pre}} + X_{\text{post}})$, where $X_{\text{pre}}$ is a RQA measure calculated based on the pre-ictal RP, and $X_{\text{post}}$ is the same RQA measure calculated based on the post-ictal RP. Note that $\Delta X$ varies between -1 and 1. Values close to -1 imply $X_{\text{pre}} \gg X_{\text{post}}$, values close to 1 imply $X_{\text{pre}} \ll X_{\text{post}}$, and values close to 0 imply $X_{\text{pre}} \approx X_{\text{post}}$.

### 2.5. Reduction in the number of configurations

Thus far, our method comprised multiple methodological choices (i.e., different frequency bands, functional network measures, network distance measures, and threshold values), which may yield redundant RPs and consequently redundant RQA results. To avoid such redundant information and inefficient computations, we reduced the number of methodological choices as follows.

In particular, for the MEG data set, we considered four frequency bands, two functional network measures, six network distance measures, and three threshold values. Different combinations of these four factors yield different RPs. However, we observe that a majority of these RPs may be redundant. Therefore, we selected representative RPs as follows. For clarity, we define a configuration as one combination of frequency band, functional network, and distance measure; we will separately consider the threshold. For example, the combination of the alpha frequency band, PLI, and the Frobenius norm is a configuration. We assessed whether some of the $4 \times 2 \times 6 = 48$ configurations yielded redundant RPs by comparing RQA results obtained using different configurations. We studied the relations among the 48 configurations for three randomly selected healthy controls from the MEG data set, using Pearson's correlation between RQA values across configurations (see Supplementary Material S5 for details). This investigation yielded a few representative configurations whose RPs we used for the subsequent analysis. We carried out this assessment of configurations independently for each of the three threshold values because, as mentioned, RPs are threshold-dependent and comparisons between RPs with different thresholds are not meaningful.

## 2.6. Statistical methods

We used the Mann-Whitney *U* test to assess whether the median of each RQA measure was different between the epilepsy and control groups in the MEG data set. We applied Bonferroni-Holm correction to correct p-values due to multiple comparisons across different configurations. We considered that the four configurations provided a family of four hypotheses for which we accounted the familywise error rate by correcting the p-values in the family. We did not correct p-values across the RQA measures because we considered that these tests corresponded to different families of hypotheses.

To evaluate whether RQA measures significantly changed from pre-ictal to post-ictal epochs in the sEEG data set, we used the Wilcoxon signed-rank test.

## 2.7. Classification methods

We further used RQA measures to classify individuals as to whether they had epilepsy in the MEG data set. We employed MATLAB's Classification Learner Toolbox which comprises a suite of 24 different classifiers, including logistic regression, trees, *k* nearest neighbor (kNN), among others. We tested the capability of different RQA measures and different configurations to classify the two groups of people. For each test we used all 24 classifiers in MATLAB's toolbox and chose the one with highest accuracy. We further tested whether combining all RQA measures from the four configurations yielded a higher accuracy. Because in this case we would be attempting a classification of 52 individuals using a relatively large number of features (44 features from 4 configurations × 11 RQA measures), we first reduced the number of features by using principal component analysis (PCA). Not to bias the principal components towards RQA measures with higher variances, we normalized the features. To avoid overfitting, we employed a 50-fold cross-validation procedure in all these classifications, which is a feature of the MATLAB's toolbox. The cross-validation consisted in partitioning the 52 individuals into 48 groups of 1 individual and 2 groups of 2 individuals, and then training the classifiers with 49 groups and testing them using the remaining group. We repeated this training-and-test procedure 50 times such that each group was used just once for testing.

## 3. Results

We applied our methods to two different data sets, a MEG and a sEEG data set. Our purpose was to test the usefulness of the methods in the assessment of dFNs in different contexts, enabling us to explore different strengths of the methods and allowing us to test a number of hypotheses in each data set.

### 3.1. Dynamic MEG functional networks

We studied the dynamics of functional networks inferred from resting-state MEG data and tested whether dFNs from people with epilepsy differ from healthy controls. We considered signals filtered in four frequency bands, two functional connectivity measures (i.e., AEC and PLI), and six distance measures between pairs of networks.

We defined a configuration as a combination of a frequency band, connectivity measure, and distance measure. We first studied the relations between the different configurations and observed that four configurations were sufficiently representative of all the configurations (see Supplementary Material S5). All of these four configurations were in the theta band. Three of them used the AEC as connectivity measure, whereas the other one used the PLI. The three representative distance measures identified together with the AEC were the Frobenius norm ($d_F$, Eq. (3)), the spectral norm ($d_S$, Eq. (4)), and the Euclidean norm between the Fiedler vectors ($d_{EF}$, Eq. (5)). The representative distance measure identified together with the PLI was the spectral norm. We denote these four configurations by AEC+$d_F$, AEC+$d_S$, AEC+$d_{EF}$, and PLI+$d_S$.

We used our recurrence analysis framework to compare the dynamics of resting-state MEG functional networks between people with epilepsy and healthy controls. First, we considered the AEC+$d_F$ configuration. For each individual, we obtained a sequence of 506 functional networks using AEC as connectivity measure (Fig. 1(b)). We then computed the distance between each pair of networks using $d_F$ as distance measure, obtaining a distance matrix (Fig. 1(c)). Next, we identified a threshold such that 5% of the points in the distance matrix apart from the main diagonal had a distance smaller than the threshold. By thresholding the distance matrix, we obtained an RP (Fig. 1(d)). Figures 2(a) and 2(b) show RPs from a healthy individual and an individual with epilepsy, respectively. We then used the 11 RQA measures to compare the RPs between the healthy and epilepsy groups. For example, Fig. 2(c) compares the trapping time ($TT$), an RQA measure, between the two groups using the AEC+$d_F$ configuration. We then tested whether the median of each RQA measure was different between the two groups using the Mann-Whitney $U$ test. We found that the recurrence time of first type ($T1$) and the recurrence time of second type ($T2$) were smaller in people with epilepsy than in controls in the AEC+$d_F$ configuration (see Figs. 2(g) and 2(k)).

We repeated the same analysis for the other three configurations. Overall, we observed smaller $T1$ and $T2$ in people with epilepsy than in controls in the AEC+$d_F$ and AEC+$d_S$ configurations; $T1$ was higher in people with epilepsy than controls in the AEC+$d_{EF}$ configuration; $TT$ was smaller in people with epilepsy than controls in the PLI+$d_S$ configuration (see Fig. 2(f)). All other differences between the two groups were not statistically significant ($p$-values were corrected with the Bonferroni-Holm procedure). Finally, we repeated the same analysis using RPs with densities of recurrence points of 1% and 10% and found similar results. These results suggest that dFNs from people with epilepsy recur more often but are less likely to be trapped into specific FNs than dFNs from healthy people.

The group differences observed above suggest that RQA measures may be used to classify individuals as to whether they had epilepsy. To confirm and quantify the predictive power of RQA measures, we performed a classification analysis. First, we employed the $T2$ measure to predict whether RPs from the AEC+$d_S$ configuration were obtained from people with epilepsy or from healthy controls. Note that this was the combination of RQA measure and configuration for which the p-value was the smallest among all combinations of an RQA measure and configuration when comparing the epilepsy and healthy groups. We performed the receiver operating characteristic (ROC) analysis and found an AUC of $0.76$, sensitivity of $0.58$, and specificity of $0.88$ (see Fig. 3(a)). We then used MATLAB's Classification Learner Toolbox to classify the two groups using the $T2$ values in the AEC+$d_S$ configuration. We found an accuracy of 69.2% using a logistic regression, i.e. 36 out of 52 individuals were correctly classified (see the blue bar in Fig. 3(b)). Next, we used the 11 RQA measures altogether and found that the classification accuracy was similar across the four configurations, ranging from 65.4 to 71.2% classification accuracy (see the black bars in Fig. 3(b)). Finally, we tested whether combining all RQA measures from the four configurations yielded a higher accuracy. To this end, as described in section 2.7, to do it we first reduced the number of features by using PCA. We observed that the first 12 principal components explained 85% of the variance of all RQA measures and all configurations. Therefore, we used them to perform the classification. These principal components yielded a slightly lower accuracy to that for the other classification methods (see red bar in Fig. 3(b)). Figure 3(b) shows the classification accuracy obtained from the best classifiers using MATLAB's Classification Learner Toolbox. The selected classifiers were the medium kNN for the AEC+$d_S$ configuration, the cosine kNN for the AEC+$d_{EF}$ configuration, the coarse tree for the AEC+$d_F$ configuration and PCA, and the fine kNN for the PLI+$d_S$ configuration. Table S1 shows the classification accuracy obtained by each classifier employed by MATLAB's Classification Learner Toolbox.

We also performed the same classification analysis by (i) using the weighted mean degree of the functional networks and (ii) by applying RQA to traditional RPs obtained from the MEG time series, rather than from the dFNs (see Supplementary Material S7 and S8). We obtained a classification accuracy of 67.3% using the weighted mean degree and an accuracy of 69.2% using the traditional RQA analysis.

### 3.2. Dynamic sEEG functional networks

We also applied our framework to dFNs inferred from sEEG data from people with drug-resistant focal epilepsy. In contrast to the resting-state MEG data, the sEEG data contained electrographic seizures. Here our purpose was to observe how dFNs change upon seizure onset and offset, whether dFNs are consistent across different seizures, and whether seizure onset may be identifiable using our framework.

We considered one broad frequency band, the Pearson's correlation as connectivity measure, and the same six distance measures between functional networks as in the study of the MEG data set. As a result of applying the reduction of configurations (see Supplementary Material S5), we focused our analysis on three distance measures: $d_F$, $d_S$, and $d_{EF}$.

We analyzed sEEG data from 10 individuals, each of them with two to four peri-ictal epochs. There were 29 peri-ictal epochs in total. Each peri-ictal epoch contained five minutes of data before seizure onset (pre-ictal), the seizure (ictal), and five minutes of data after seizure offset (post-ictal). Our aim was to observe whether functional networks show stereotypical dynamics throughout seizures, whether we can detect seizures, and how the pre-ictal, ictal, and post-ictal networks relate to each other.

For each peri-ictal epoch of each individual, we computed a sequence of $M$ functional networks using Pearson's correlation, where $M$ varied between 1542 and 1822 with median 1632, depending on the length of the ictal periods. Below, we present results obtained using the Frobenius norm, Eq. (3), to compute the distance between functional networks, and using a threshold such that the RPs had a density of recurrence points equal to $0.05$. We obtained similar results using the two other distance measures and using the two other threshold values (i.e. thresholds such that the RPs had a density of recurrence points equal to $0.01$ and $0.1$).

### 3.2.1. *RPs of multiple peri-ictal epochs*

To assess whether similar functional networks are involved in different pre-ictal, ictal and post-ictal epochs from different peri-ictal epochs of an individual, we first concatenated the sequences of functional networks from different peri-ictal epochs of the individual. Next, we computed the distance between each pair of networks in the concatenated sequence to obtain a distance matrix (see Fig. 1(c)). Then, by thresholding the distance matrix, we obtained an RP (see Fig. 1(d)). Figures 4(a) and 4(b) show RPs for two individuals. Each RP contains three peri-ictal epochs. We observe that most of the recurrence points are located in the pre-ictal periods (i.e., in the first, fourth, and seventh diagonal blocks) and that there is also a high density of recurrence points in off-diagonal blocks corresponding to the cross relation between different pre-ictal periods. The ictal periods also show high density of recurrence points within the same peri-ictal epoch and between different peri-ictal epochs. In contrast, the post-ictal periods show a low density of recurrence points both within and between epochs. These results imply that pre-ictal and ictal functional networks are more similar between themselves than post-ictal functional networks. There is also a considerable frequency of recurrence between pre-ictal and ictal networks, both within and between epochs. This result suggests that the functional networks involved in seizures emerge before seizure onset. In contrast, the functional networks after seizure offset are relatively different from networks during both pre-ictal and ictal periods.

To quantify the observations made from Figs. 4(a) and 4(b), we measured the relative recurrence rate ($RR$) defined as the fraction of the actual recurrence points in a block divided by the fraction of recurrence points expected by chance, i.e., 5% (because the overall density of recurrence points was set to 5%). Figure 4(c) shows the relative $RR$ of each type of block in the RPs for all individuals. The two leftmost bars in each type of block in Fig. 4(c) correspond to the two individuals whose RPs are shown in Figs. 4(a) and 4(b) and confirm our previous observations. The relative $RR$ of these two individuals further shows that, whereas the recurrence rate between a pre-ictal period and an ictal period belonging to different peri-ictal epochs is not higher than chance in Fig. 4(a), it is higher than chance in Fig. 4(b). Overall, Fig. 4(c) shows that the pre-ictal periods have recurrence rates higher than chance in 9 out of the 10 individuals. The ictal periods also have higher recurrence rates than chance in all but one individual. Post-ictal periods have low recurrence rates, except in two individuals. The cross relations between different pre-ictal periods, between pre-ictal and ictal periods, and between ictal periods show considerable variability in terms of the relative $RR$, which is either below or above chance depending on the individual.

In contrast, the cross relations between pre-ictal and post-ictal periods, between ictal and post-ictal periods, and between different post-ictal periods is consistently lower than chance, except in one individual that has high relative $RR$ in the cross relation between different post-ictal periods. These results support that the observations made from Figs. 4(a) and 4(b) for two individuals generalize to most of the 10 individuals. The individual represented by the third bar from the left in Fig. 4(c) (shown in dark green) is an outlier.

To test whether these results were confounded by the different number of peri-ictal epochs per individual, we further computed the relative $RR$ using only two peri-ictal epochs from each individual. For individuals with more than two peri-ictal epochs, we computed the relative $RR$ for all combinations of two peri-ictal epochs. Figure S3 shows the result of this analysis. We observe that the results in Fig. S3 are similar to those in Fig. 4(c), meaning that most combinations of two peri-ictal epochs from each individual produce approximately the same relative $RR$ pattern as that produced by all peri-ictal epochs combined together. Thus, these results suggest that the results in Fig. 4(c) are not confounded by the number of peri-ictal epochs of each individual.

### 3.2.2. RPs of single peri-ictal epochs

We then assessed whether or not RQA may be able to detect seizures. RPs comprising multiple peri-ictal epochs are not appropriate for this purpose, because a peri-ictal epoch may have a disproportionately high or low fraction of recurrence points compared to other peri-ictal epochs for the same individual. Therefore, here we did not concatenate the sequences of functional networks over different peri-ictal epochs. Instead, we constructed an RP for each of the 29 peri-ictal epochs.

Figures 5(a) and 5(b) show RPs from a peri-ictal epoch of different individuals. These figures show that the pre-ictal periods present a higher recurrence rate than both ictal and post-ictal periods, in agreement with Fig. 4. Transitions in the RPs are also noticeable at the seizure onsets. Furthermore, whereas there is recurrence of pre-ictal networks in the ictal period, there is neither recurrence of pre-ictal nor ictal networks in the post-ictal period. Additionally, the post-ictal period has a small number of recurrence points.

To assess the potential of RQA to detect seizure onset, we computed the $\tau$-recurrence rate ($RR_\tau$) for the RPs shown in Figs. 5(a) and 5(b). The $\tau$-recurrence rate quantifies the frequency of recurrence when pairs of functional networks are $\tau$ time points apart. When $\tau$ is larger than the seizure onset time, $RR_\tau$ measures recurrences of pre-ictal networks in the ictal and post-ictal periods, as well as recurrences of ictal networks in the post-ictal period. When $\tau$ is larger than the seizure offset time, $RR_\tau$ only uses recurrences of pre-ictal networks in the post-ictal period. In contrast, when $\tau$ is smaller than the seizure onset time, $RR_\tau$ also accounts for recurrences within the pre-ictal period and within the post-ictal period, as well as within the ictal period when $\tau$ is also smaller than the duration of the seizure. Because the relative $RR$ is large in the pre-ictal period, but small in the cross relation between pre- and post-ictal periods, we expect that $RR_\tau$ is large if and only if $\tau$ is smaller than the seizure onset time. Furthermore, we expect that $RR_\tau$ decreases as $\tau$ increases from zero towards the seizure onset time because the contribution of recurrences within the pre-ictal period to $RR_\tau$ decreases as $\tau$ increases in this range. Therefore, $RR_\tau$ may be a useful tool for detecting seizure onset.

The $\tau$-recurrence values for the two individuals are shown in Figs. 5(c) and 5(d). As anticipated, the $RR_\tau$ decreases towards 0 as $\tau$ approaches the value corresponding to the seizure onset time in Fig. 5(c) and corresponding to the seizure offset time in Fig. 5(d). For both individuals, $RR_\tau$ remains near zero for larger $\tau$ values. The difference between the two individuals results from the fact that pre-ictal networks do not often recur in the ictal period for the individual represented in Figs. 5(a) and 5(c), but they do recur for the individual represented in Figs. 5(b) and 5(d). Figures 5(c) and 5(d) also display the results for other peri-ictal epochs from the same individuals, confirming qualitatively the same observations. We also plotted the mean and standard deviation of $RR_\tau^{\text{null}}$ above and below its mean as the shaded regions, which represents a distribution of $RR_\tau$ obtained from randomized RPs. The mean of $RR_\tau^{\text{null}}$ is 0.05 for any $\tau$, which is the density of recurrence points in the RP. This is because, in a randomized RP, every diagonal line, corresponding to a value of $\tau$, has on average the same density of points as the original RP. Figures 5(c) and 5(d) indicate that $RR_\tau$ is significantly larger than for the randomized RPs when $\tau$ is small and that it is significantly closer to zero than the randomized RPs as $\tau$ becomes large and approaches the seizure onset.

We performed the same analysis for the other 8 individuals and found that $RR_\tau$ decreases to values close to zero at seizure onset or close to seizure offset in the majority of the peri-ictal epochs considered (see Fig. 6). There are three notable exceptions (the blue line in Fig. 6(d), and the green lines in Figs. 6(f) and 6(g)), which do not show a decrease in $RR_\tau$ at either seizure onset or offset. Overall, the results suggest that a decrease in $RR_\tau$ to values close to zero at a certain $\tau$ value is indicative of a seizure onset or offset at time $\tau$.

### 3.2.3. *RPs of individual pre- and post-ictal periods*

We have shown that the recurrence rate within the pre-ictal period is much higher than within the post-ictal period (see Fig. 4(c)). An RP for a peri-ictal epoch, which by definition contains a pre-ictal, ictal, and post-ictal period altogether (e.g., Fig. 5), does not allow us to compare recurrence features of functional networks within the pre-ictal period to those within the post-ictal period due to the difference in density of recurrence points in the two periods. To address this limitation, and to compare the dynamics of functional networks between pre-ictal and post-ictal epochs, we constructed RPs containing only a pre-ictal or a post-ictal period. We did not consider RPs only containing an ictal period because the ictal periods were typically too short for RQA. Figures 7(a) and 7(b) show an RP from a pre-ictal and post-ictal period, respectively, belonging to the same peri-ictal epoch. By construction, the two RPs had the same recurrence rate, which allowed us to compare other properties of the RPs between the pre-ictal and post-ictal periods.

Figures 7(c)-(e) show the variation of $DET$, $\langle L \rangle$, and $L_{\max}$ between all pre- and post-ictal RPs within the same peri-ictal epoch for all individuals. We observe that in most peri-ictal epochs these three measures are larger in the post-ictal than pre-ictal

periods (i.e., positive variation values), meaning that post-ictal RPs have typically longer diagonal lines than pre-ictal RPs. We used the Wilcoxon signed-rank test to assess the significance of these changes and found $\Delta DET$ and $\Delta L_{\max}$ significant and $\Delta \langle L \rangle$ at the boundary of significance. This result suggests that post-ictal dFNs may be more deterministic than pre-ictal dFNs. For some individuals, this variation is consistently positive (i.e., the post-ictal values are larger than the pre-ictal values) for all their peri-ictal epochs (e.g., the individual represented by the right-most triangles in Figs. 7(c)-(e)).

Figure S4 shows the variation of the other RQA measures. The $ENTR$, $LAM$, $TT$, $V_{\max}$, and $Trans$ values tended to be larger in the post-ictal than pre-ictal periods, whereas the opposite was the case for $T1$, $T2$, and $RTE$. The measures based on vertical lines ($LAM$, $TT$ and $V_{\max}$) indicate that dFNs are more likely to be trapped in slowly changing functional networks in the post-ictal than the pre-ictal period.

4. **Discussion**

In the present study, we have proposed to use RPs and RQA to study dFNs. The framework consists of assessing the distance between functional networks at different times and define recurrences if the distance is within a threshold. By doing so, one obtains an RP on which one can perform RQA. The RQA measures inform us about the underlying dynamics of the functional networks. We applied this framework to two data sets, (i) resting-state MEG recordings from people with JME and healthy controls, and (ii) sEEG recordings containing peri-ictal epochs of people with drug-resistant focal epilepsy undergoing pre-surgical evaluation. The purpose of these two independent analyses was to show the broad applicability of the framework and its potential to address a wide range of hypotheses. In the MEG data set, we found that RQA measures for dynamic resting-state functional networks differed between people with epilepsy and healthy controls. In the sEEG data set, we found that pre-ictal functional networks show high recurrence not only within pre-ictal periods but also between different pre-ictal periods, and that they also recur during ictal periods. This result implies not only that functional networks involved in seizures emerge before seizure onset, but also that they are consistent across different seizures. We also observed that RQA measures were capable of detecting seizures. Finally, we observed that post-ictal dFNs are typically more deterministic and more likely to be trapped into certain networks compared to pre-ictal dFNs.

Using the MEG data set, we found significantly smaller recurrence time in people with epilepsy compared to healthy controls (for both first ($T1$) and second type ($T2$) and in both the AEC+$d_F$ and AEC+$d_S$ configurations; see Figs. 2(g), 2(h), 2(k) and 2(l)). This result implies that AEC functional networks recur more quickly in people with epilepsy than in controls. This finding suggests that the space of possible functional networks may be more limited in epilepsy. We speculate that a reduced repertoire of functional networks may lead to functional brain deficits. This is in agreement with the fact that cognitive deficits are commonly observed in people with epilepsy (Holmes, 2015). Future work should test the hypothesis that reduced recurrence times predict cognitive deficits. One should also further examine such findings across different configurations (see Supplementary Material S9).

We then used RQA measures to classify people as to whether they had epilepsy and found an accuracy of 71.2% (see Fig. 3). This classification power is similar to that achieved with single resting-state scalp EEG functional networks for people with idiopathic generalized epilepsy (Schmidt et al., 2016). In our study, we considered juvenile myoclonic epilepsy, which is a specific type of generalized epilepsy. Future work should test whether our findings extend to other types of both generalized and focal epilepsy. Furthermore, drug-naïve individuals should be recruited in order to exclude the possibility that the differentiation between people with epilepsy and healthy controls is due to medication intake, and not epilepsy. Notably, the opportunity to use resting-state data to distinguish people with epilepsy from healthy people is of great clinical value because current clinical practice depends upon the observation of epileptiform discharges which are not always apparent in a scalp EEG session (Smith, 2005). Since scalp EEG is more inexpensive and available than MEG, future work should also test whether our findings generalize to dFNs inferred from scalp EEG.

We also assessed whether the combination of our approach with others achieved a superior classification accuracy. The mean number of connections in EEG functional networks has been used to differentiate between people with idiopathic generalized epilepsy and controls (Chowdhury et al., 2016). Therefore, we computed the weighted mean degree of our MEG functional networks and used it to classify the individuals (see Supplementary Material S7). We obtained 67.3% accuracy, which was slightly lower than the classification accuracy that we attained with the RQA measures on dFNs. The accuracy did not improve by combining the weighted mean degree with the RQA measures. Additionally, we also used traditional RQA measures applied to RPs obtained from the MEG time series to classify the individuals (see Supplementary Material S8). We found an accuracy of 69.2%, i.e., similar to the accuracy achieved using the RQA for dFNs. Again, we found that combining the two types of RQA did not improve the accuracy of the classification. This result suggests that the two types of analysis may extract similar information. However, this does not imply that the two methods are equivalent. By inferring dFNs and studying their recurrences, we are examining the recurrences of statistical dependencies between the MEG signals. In contrast, the traditional recurrence analysis looks at the recurrence of the MEG signals themselves. Another difference between the two methods is that by computing dFNs we reduced the time resolution of the recurrence analysis, i.e., from 4 ms in the traditional RPs to 400 ms in our RPs. Future work should further explore the relation between the two methods and test whether they may complement each other on various data sets.

We highlight that our analysis in the MEG data set was based on functional networks inferred in the theta band, and that their recurrence was representative of the recurrence of dFNs in other frequency bands (see Supplementary Material S5). This is an unexpected result given that MEG functional networks are usually different across different frequency bands (Tewarie et al., 2016). Notwithstanding, our results suggest that the recurrence of dFNs, at least using measures afforded by RQA, may be similar across frequency bands. Further investigation of the recurrence of dFNs in different frequency bands is needed to ascertain their cross-frequency relations.

Using the sEEG data set, we found considerable recurrence of pre-ictal and ictal functional networks in their respective periods (see Fig. 4(c)). We further observed that ictal functional networks recurred in different seizures for most of the individuals, and that there was considerable recurrence of functional networks between different pre-ictal periods and between pre-ictal and ictal periods. Such information may be useful when evaluating whether there are one or more leading networks that sustain seizures. Non-recurrence of functional networks across different seizures may be a contraindication for epilepsy surgery because multiple seizure focus may be involved. In contrast, functional networks that recur both in pre-ictal and ictal periods across different peri-ictal epochs may support epilepsy surgery and they may be used to inform where to perform the resection (Goodfellow et al., 2016; Lopes et al., 2017; Lopes et al., 2018; Kini et al., 2019). Future work should test whether the combination of our framework with other recent methods that use functional networks to inform epilepsy surgery and predict surgery outcome (Goodfellow et al., 2016; Lopes et al., 2017; Lopes et al., 2018; Kini et al., 2019) yield superior predictions.

Seizure detection is highly relevant not only for seizure management (Jory et al., 2016), but also to assist neurologists in the analysis of EEG (Adeli et al., 2003). We observed that the $\tau$-recurrence rate ($RR_\tau$) values decreased to close to zero when $\tau$ approached the seizure onset or offset time in peri-ictal epochs from all individuals, except for 3 peri-ictal epochs (see Figs. 5(c), 5(d), and 6). $RR_\tau$ takes advantage of the fact that functional networks frequently recur within the pre-ictal period and relatively rarely within the post-ictal period. These findings suggest that our framework may be useful for seizure detection. Future work should assess whether the decrease in $RR_\tau$ is specific to seizures, which one can assess by additionally measuring $RR_\tau$ far from seizures (i.e., in the inter-ictal periods, which we did not consider in this study because of lack of such data). In such future work, one should consider assessing $RR_\tau$ not relative to randomly shuffled RPs as we did in Figs. 5(c), 5(d), and 6, but rather relative to an inter-ictal baseline. Additionally, our methods should be compared to other recent methods to detect seizures (Leijten et al., 2018). Beyond sEEG data, it should be tested whether this method may be useful for seizure detection from scalp EEG.

We also found that the dynamics of functional networks tended to be more deterministic and more frequently trapped in certain functional networks in the post-ictal period than in the pre-ictal period (see Figs. 7 and S4). We hypothesize that individuals that show consistent differences in RQA measures between pre-ictal and post-ictal periods across seizures may be particularly suited to receiving neurostimulation treatment (Morrell, 2006). Neurostimulation can be used to modulate brain activity of people with epilepsy to avoid the emergence of seizures (Morrell, 2006). Thus, we suggest that our framework may be used for finding whether an individual presents differences in dFNs between the pre-ictal and post-ictal periods that are consistent across seizures. Such a consistency supports the use of a single stimulation strategy that may be reliable across peri-ictal epochs. Furthermore, our framework also informs us of the specific differences between pre-ictal and post-ictal dFNs. A personalized stimulation strategy may then be designed such that it modulates the dynamics of pre-ictal functional networks into those of the post-ictal period, whereby avoiding seizures (Dalkilic, 2017).

There is a number of confounding factors to consider when assessing our sEEG results. First and foremost, each individual had different causes of epilepsy with seizures emerging from different brain regions and each individual received different electrode implantations. Consequently, functional networks from different individuals had different numbers of nodes (i.e. channels) and covered different regions of the brain. Such differences may account for some of the variability observed among individuals. Second, even for the same individual, different peri-ictal epochs contained seizures of varying lengths, and peri-ictal epochs were at different time distances from other seizures. The distance to other seizures may be an important confounding factor when comparing pre-ictal and post-ictal periods. Third, although we used 5 min before and after a seizure as a pre-ictal and post-ictal period, respectively, this choice was motivated by the need of a sufficient number of functional networks for our analysis, rather than a clinically motivated decision. Different definitions of pre-ictal and post-ictal periods may yield different results.

Although we applied our framework to dFNs inferred from MEG and sEEG data, in principle the framework is applicable to any kind of time-varying network or matrix. An important requirement is to have a sufficient number of networks or matrices to reliably evaluate their dynamics and recurrences (Marwan et al., 2007). Thus, the framework may not be applicable to study dFNs derived from typical fMRI experiments due to a relatively small number of time points (Hutchison et al., 2013).

For this reason, we did not explore the dynamics of functional networks during seizures in the sEEG data set; seizures were not long enough for our analysis. Additionally, although we focused on epilepsy in the present study, our framework may also be suitable to studying healthy brain function or other brain disorders.

There are other computational approaches to dFNs. Common approaches include tracking certain network measures over time (Sizemore and Bassett, 2018), using hidden Markov models (Eavani et al., 2013; Sourty et al., 2016; Vidaurre et al., 2018), and considering dynamic networks as multilayer networks (de Domenico et al., 2013; Kivelä et al., 2014; Sizemore and Bassett, 2018). Other recent approaches have used distance matrices to evaluate dFNs from fMRI (Cabral et al., 2017) and dynamic correlation matrices from scalp EEG (Rosch et al., 2018). Future work should compare these different methods to our framework to find which one better characterizes dFNs in epilepsy and other contexts, and assess whether these approaches complement each other.

In conclusion, we propose a new framework to assess dFNs based on recurrence analysis. We applied the framework to source-localized resting-state MEG data and found that it is capable of distinguishing people with epilepsy from healthy controls. We also used the framework to assess sEEG dFNs and found supporting evidence that it may be useful for seizure detection. The framework further opens avenues to test new hypotheses, namely, to advance methods of epilepsy surgery assessment, and to potentially inform neurostimulation strategies. The framework may also be used to study dFNs in healthy brain functions and in other neurological disorders.

**Acknowledgments**

ML, JZ, LL, and NM gratefully acknowledge funding from the GW4 Accelerator Fund. JZ further acknowledges the financial support of the European Research Council [grant number 716321]. DK was supported by an EPSRC PhD studentship [grant number EP/N509449/1]. KH acknowledges support from the Health and Care Research Wales: Clinical Research Time Award and the Wales BRAIN Unit. QC is supported by grants from Natural Science Foundation of China (31871138), and by the Chang Jiang Scholars Program (2016). LL gratefully acknowledges partial support of the Canada Research Chairs program. NM acknowledges support from AFOSR European Office [grant number FA9550-19-1-7024]. We thank Victor Küpper for having contributed for the initial development of the methodology presented in this manuscript.

**Conflict of Interest**

None

**Author Contributions**

Marinho A. Lopes: Conceptualization; Methodology; Software; Formal analysis; Writing - Original Draft; Writing - Review & Editing; Visualization.

Jiaxiang Zhang: Conceptualization; Writing - Review & Editing; Project administration; Funding acquisition.

Dominik Krzemiński: Software; Formal analysis; Writing - Review & Editing.

Khalid Hamandi: Investigation; Data Curation; Writing - Review & Editing

Qi Chen: Investigation; Data Curation; Writing - Review & Editing.

Lorenzo Livi: Conceptualization; Methodology; Writing - Review & Editing; Funding acquisition.

Naoki Masuda: Conceptualization; Methodology; Writing - Original Draft; Writing - Review & Editing; Supervision; Project administration; Funding acquisition.

**Abbreviations**

dFN: dynamic functional networks; JME: juvenile myoclonic epilepsy; RP: recurrence plot; RQA: recurrence quantification analysis; sEEG: stereo EEG;

**Data Accessibility Statement**

All materials (functional networks and code) are available upon request (contact m.lopes@exeter.ac.uk).

**Figure captions**

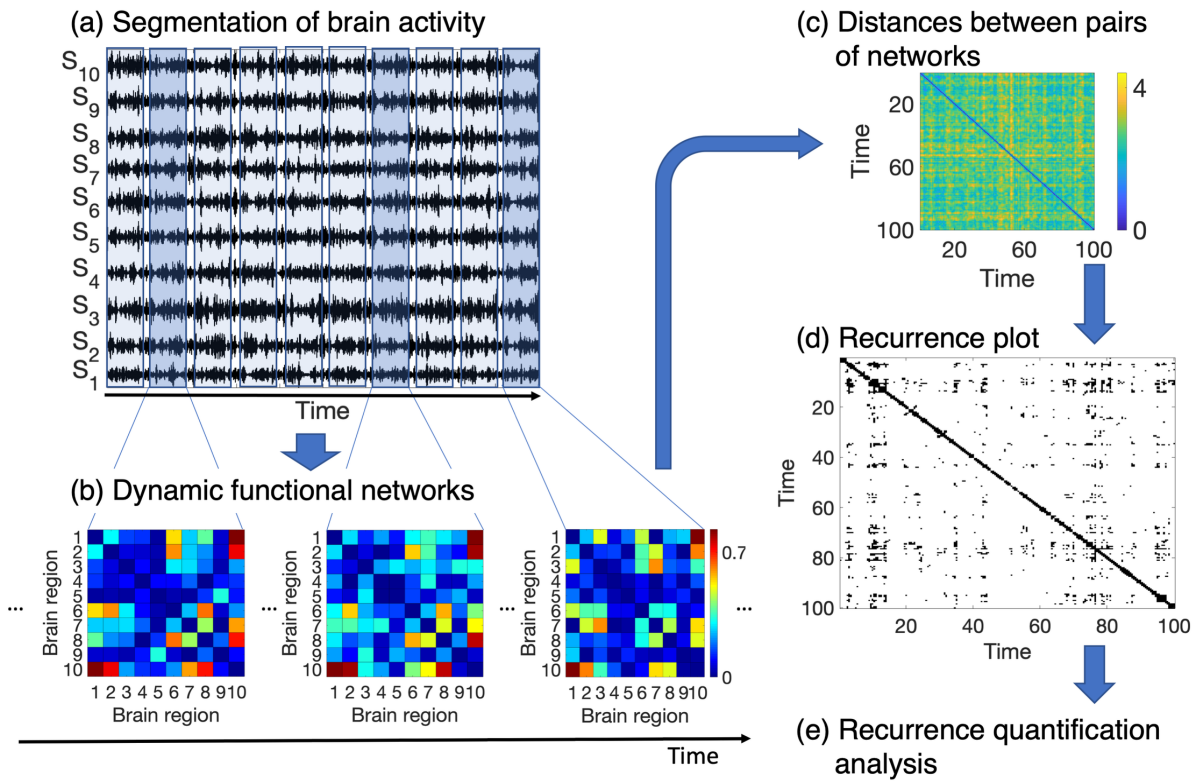

Figure 1: Scheme of the data analysis procedure to apply recurrence quantification analysis to dynamic functional brain networks. (a) We segment brain activity into windows. (b) From each window, we infer a functional brain network. (c) We employ a distance measure for assessing the similarity between functional networks at each pair of time windows, resulting in a distance matrix. (d) We obtain a recurrence plot by assessing the distances between functional networks. Black dots correspond to entries in the distance matrix shown in (c) whose value is smaller than a threshold. (e) We then use recurrence quantification analysis to interrogate the recurrence plot and consequently characterize the dynamics of the functional networks. Note that, in our analysis, the windows represented in (a) overlapped with adjacent windows. In this figure, we avoided representing overlapping windows for visual clarity.

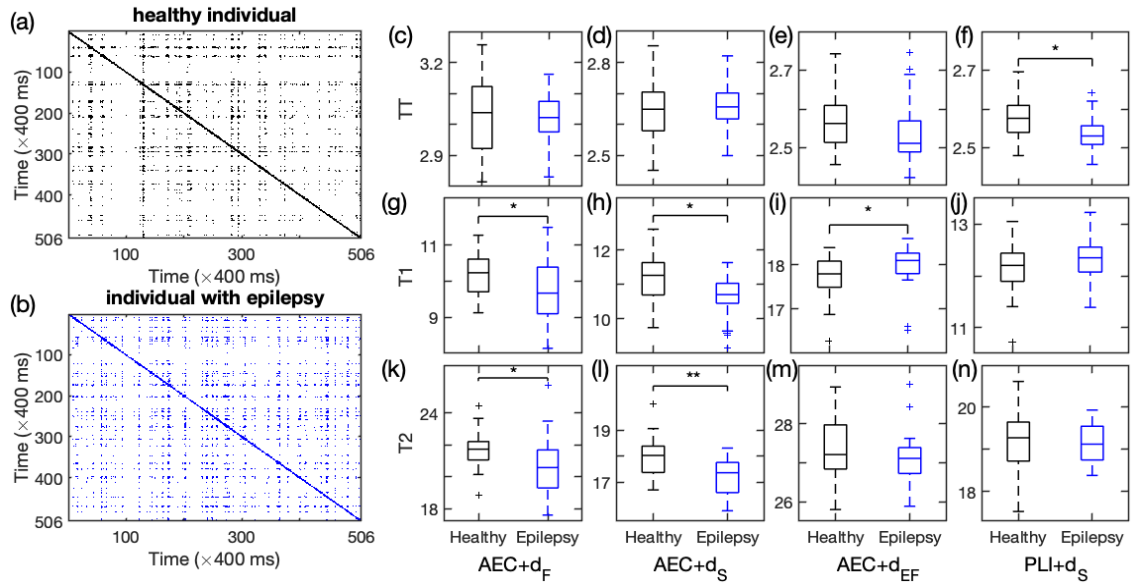

Figure 2: Recurrence quantification analysis of dynamic MEG functional networks. (a) Representative RP from a healthy individual. (b) Representative RP from an individual with epilepsy. We used the AEC+$d_F$ configuration. (c-f) Box plots of the trapping time ($TT$) of the RPs from healthy controls and people with epilepsy. Each of the four panels in this row compares controls to people with epilepsy in one of the four configurations, i.e. AEC+$d_F$, AEC+$d_S$, AEC+$d_{EF}$, and PLI+$d_S$. Similarly, the rows of panels (g-j) and (k-n) show box plots for the recurrence time of first type ($T1$) and second type ($T2$), respectively, across the four configurations. In (f), (g), (h), (i), (k) and (l), significant differences between controls and people with epilepsy are indicated by asterisks (one asterisk represents $p < 0.05$, and two asterisks $p < 0.01$, Mann-Whitney $U$ test, Bonferroni-Holm corrected). We used a density of recurrence points of $0.05$ to define the threshold $\epsilon$.

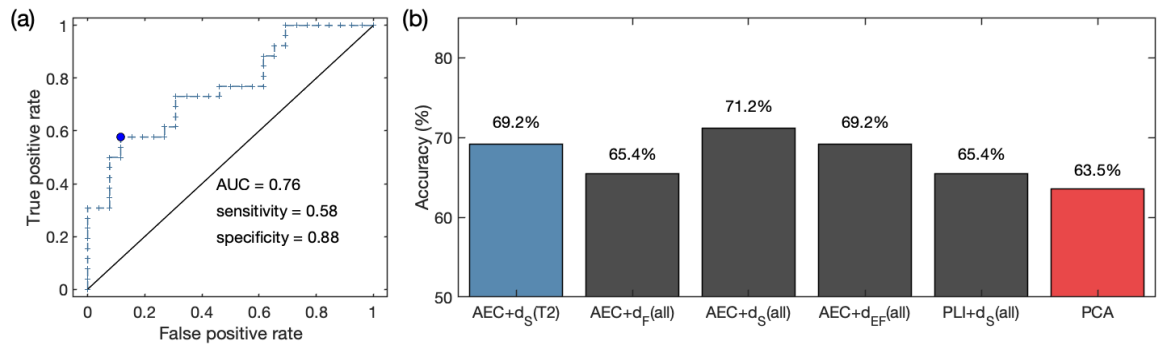

Figure 3: Classification of people with epilepsy using RQA applied to dFNs inferred from MEG data. (a) Receiver operating characteristic (ROC) analysis to classify people as either healthy or with epilepsy using the recurrence time of second type ($T2$) from the configuration AEC+$d_S$. The circle represents the optimal operating point of the ROC curve, for which the sensitivity is equal to 0.58, and the specificity is equal to 0.88. The area under the curve (AUC) is equal to 0.76. (b) Accuracy of the classification using different features from the RQA analysis and 50-fold cross-validation: the recurrence time of second type ($T2$) from the AEC+$d_S$ configuration (blue bar), all RQA measures from each of the four representative configurations (grey bars), and 12 principal components explaining 85% of the variance of all RQA measures across the four configurations (red bar).

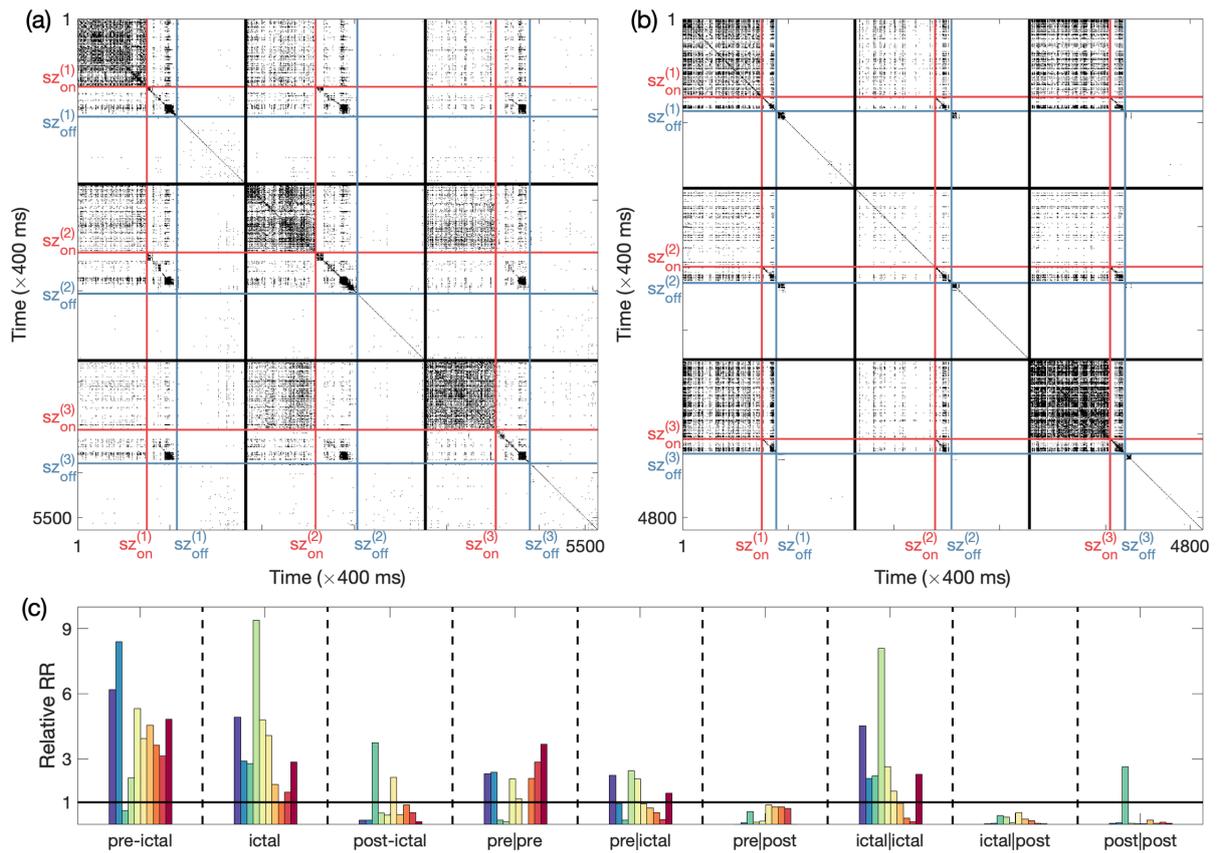

Figure 4: Recurrence plots (RPs) and relative recurrence rate ($RR$) of sEEG dFNs from individuals with epilepsy. (a) RP for three peri-ictal epochs of one individual. The thick black lines separate the three peri-ictal epochs. The thin red and blue lines indicate the seizure onset and offset in each peri-ictal epoch, respectively. Panel (b) is equivalent to (a) but for a different individual. (c) Relative $RR$ for all individuals with epilepsy. For each individual, we consider the relative $RR$ in nine types of blocks in the RPs: the label 'pre-ictal' corresponds to the three pre-ictal diagonal blocks of the RP; 'ictal' corresponds to the three ictal diagonal blocks; 'post-ictal' corresponds to the three post-ictal diagonal blocks; 'pre|pre' corresponds to the six off-diagonal blocks that compare different pre-ictal periods; 'pre|ictal' corresponds to the 18 off-diagonal blocks that compare pre-ictal and ictal periods; 'pre|post' corresponds to the 18 off-diagonal blocks that compare pre-ictal and post-ictal periods; 'ictal|ictal' corresponds to the six off-diagonal blocks that compare different ictal periods;

'ictallpost' corresponds to the 18 off-diagonal blocks that compare ictal and post-ictal periods; and the 'postlpost' corresponds to the six off-diagonal blocks that compare different post-ictal periods. For each type of block, we plot 10 bars, one for each individual. The two leftmost bars correspond to the RPs shown in (a) and (b), respectively. The chance level, i.e., 1, is represented by the solid horizontal line. Figure S2 shows that the results in (c) remain similar for other configurations (i.e. using different distance measures between networks).

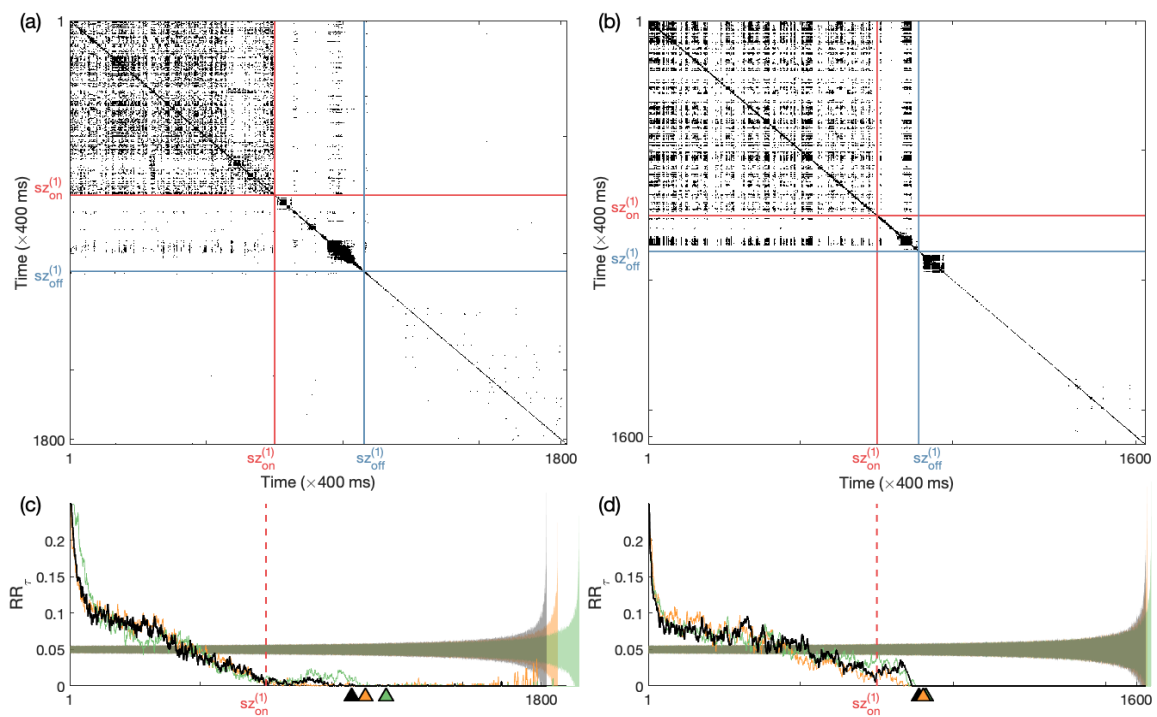

Figure 5: Recurrence plots (RPs) and $\tau$-recurrence rate ($RR_\tau$) of sEEG dFNs during single peri-ictal epochs. (a) RP of dFNs during one peri-ictal epoch of one individual with epilepsy. This peri-ictal epoch corresponds to the first peri-ictal epoch shown in Fig. 4(a). The seizure onset and offset are indicated by the red and blue lines, respectively. (b) Same as (a) but for a peri-ictal epoch of a different individual, i.e., the first peri-ictal epoch in Fig. 4(b). (c) $RR_\tau$ for three peri-ictal epochs for the individual considered in (a). (d) Same as (c) but for the RP shown in (b). In (c) and (d), the black line represents the $RR_\tau$ computed from the RP in (a) and (b), respectively; the other two lines correspond to the other two peri-ictal epochs of the same individual; the dashed lines indicate time lags equal to the time of seizure onset for all peri-ictal epochs; the triangles indicate time lags equal to the time of seizure offset, where the color of the triangle matches to that of the $RR_\tau$ curve. The shaded areas represent the standard deviation of $RR_\tau^{\text{null}}$ above and below its mean obtained from 100 random shuffles of the RPs of each peri-ictal epoch. At large $\tau$, the standard deviation of $RR_\tau^{\text{null}}$ increases, because the diagonal lines at large $\tau$ in the RP matrix have fewer elements, i.e., $M - \tau$ elements. Therefore, for large $\tau$, $RR_\tau$ is highly quantized; it can take values $0, 1/(M - \tau), 2/(M - \tau), \ldots, 1$ (see Eq. (6)), resulting in a large statistical fluctuation in $RR_\tau$.

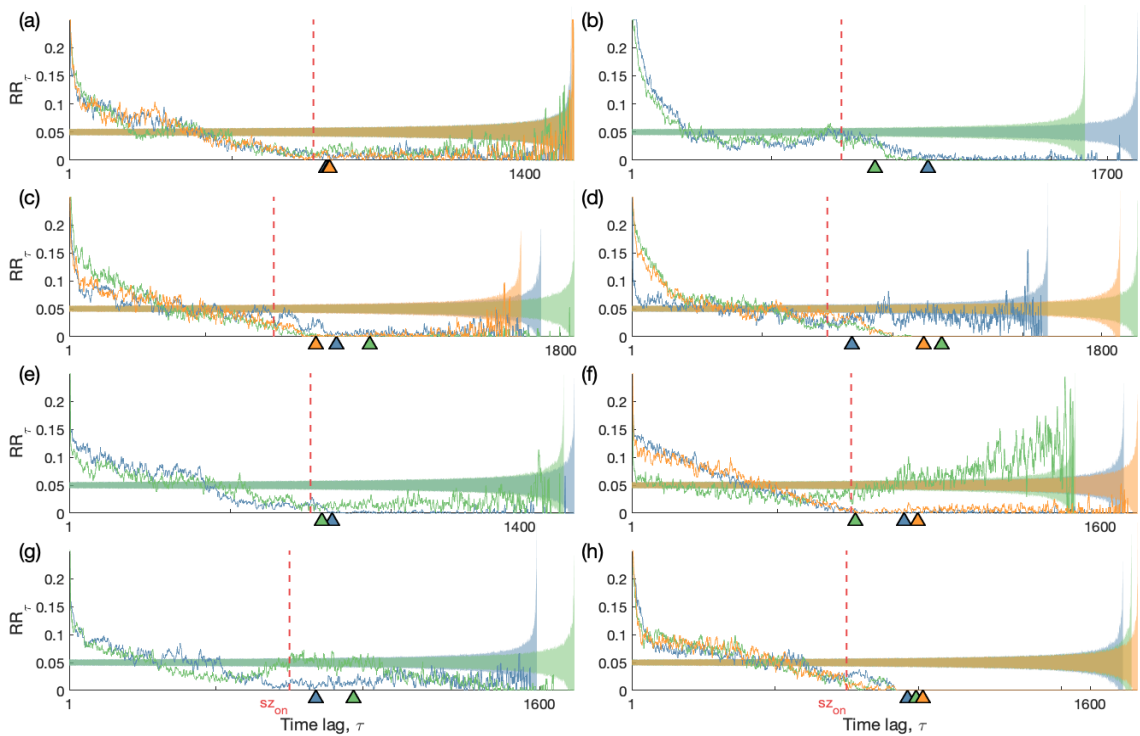

Figure 6: $\tau$-recurrence rate ($RR_\tau$) of sEEG dFNs during single peri-ictal epochs for eight individuals. Each panel represents a different individual. Each individual had between 2 and 4 peri-ictal epochs, which are represented as lines in different colors. The dashed lines indicate the time lag equal to the time of seizure onset. The colored triangles indicate the time lag equal to the time of seizure offset of the $RR_\tau$ curve in the same color. The shaded areas represent the standard deviation of $RR_\tau^{\mathrm{null}}$ above and below its mean obtained from 100 random shuffles of the RPs of each peri-ictal epoch.

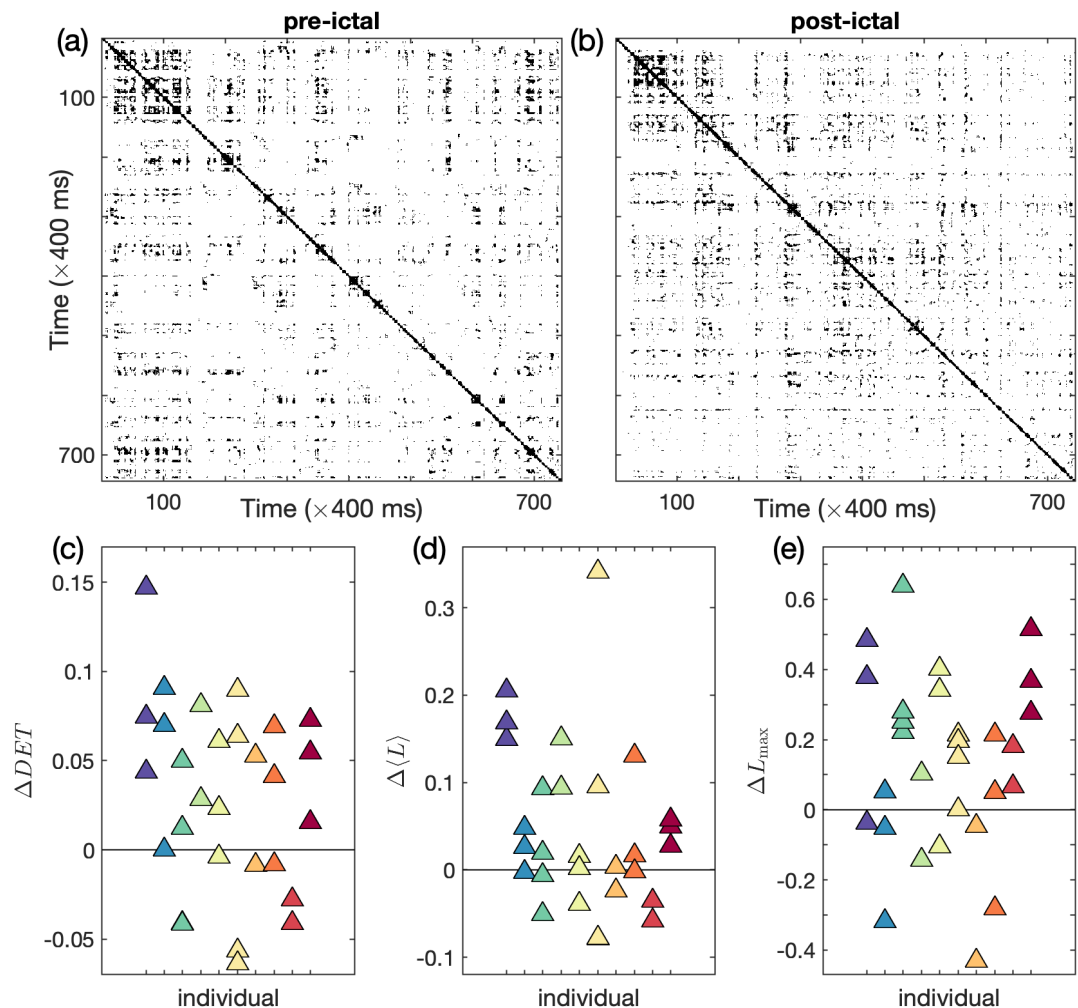

Figure 7: Comparison between pre- and post-ictal recurrence plots (RPs) of sEEG dFNs. (a) RP of dFNs during a pre-ictal period of one individual with epilepsy. (b) RP

of dFNs during a post-ictal period corresponding to the same individual and the same peri-ictal epoch as in panel (a). Panels (c), (d), and (e) show the $DET$, $\langle L \rangle$, and $L_{\max}$ variation, respectively, between pre- and post-ictal RPs. Wilcoxon signed-rank tests showed that $DET$ and $L_{\max}$ significantly changed from pre- to post-ictal epochs, and the variation of $\langle L \rangle$ was at the boundary of significance (the p-values were 0.01 for $\Delta DET$ and $\Delta L_{\max}$, and 0.05 for $\Delta \langle L \rangle$). A triangle corresponds to a peri-ictal epoch. The color of the triangles as well as their horizontal positions distinguishes the individuals.

**Supporting Information**

**Text S1: Overlap between consecutive functional networks**

Here we describe how we determined the overlap between consecutive segments to infer dFNs. We determined the amount of overlap as a compromise between two considerations; first, the larger the overlap, the larger is the number of functional networks that are inferred from an epoch of data, and in turn the larger and potentially richer is the resulting RP; second, an excessively large overlap produces trivial recurrences between functional networks adjacent in time that are not of interest (see section 2.4).

For a healthy control randomly chosen from the MEG data set, we first computed time-varying functional networks from the MEG signals filtered in the theta band using AEC (see section 2.3 and Supplementary Material S2). We divided the data into segments with maximal overlap. In other words, consecutive segments of 500 samples shared 499 samples (i.e., sliding the window only one sample from one segment to the next). This procedure yielded a sequence of functional networks $A(t) = (a_{ij}(t))$, where $t = 1, ..., T$ and $T$ was the number of networks, i.e., the number of segments. We then employed the spectral norm distance measure, Eq. (5), to compute the distance between each pair of networks and hence a distance matrix (see Fig. S1(a)). Note that elements close to the main diagonal of the distance matrix correspond to large overlaps between pairs of functional networks, whereas elements far from the main diagonal correspond to small or zero overlaps. By definition, all main diagonal elements of this matrix are equal to zero because they correspond to pairs of identical networks. Elements close to the main diagonal were close to zero, whereas elements far from the main diagonal oscillated between different distance values. Therefore, the present RP would show a high density of recurrence points close to the main diagonal. Such recurrence points correspond to the so-called *tangential motion* and would not provide information about actual recurrences in the dynamical system (Marwan et al., 2007). Therefore, in RPs that we use to perform RQA, we should choose an overlap small enough to be able to avoid these meaningless recurrence points, but simultaneously large enough not to miss information about the temporal evolution of the functional networks.

To this end, we plotted the distances between pairs of functional networks against the amount of the overlap in Fig. S1(b). All the $T - 1$ pairs of networks $(A(t_a), A(t_b))$ with $|t_a - t_b| = 1$ corresponded to the maximal overlap (499/500). The distance between each of these 499 pairs of networks is represented as a dot located at the overlap value of $499/500 \times 100\% = 99.8\%$ in Fig. S1(b). All the $T - 2$ pairs of networks with $|t_a - t_b| = 2$ corresponded to an overlap of $498/500 \times 100\% = 99.6\%$. In general, all pairs of networks with $|t_a - t_b| = n$ and $n < 500$ corresponded to an overlap of $(500 - n)/500 \times 100\%$. (For $n \geq 500$ the overlap was zero.) Figure S1(b) suggests that an 80% overlap between consecutive segments represents a good compromise because the minimum distance between pairs of networks, which would be relevant to recurrence events, does not substantially increase if we further decrease the overlap. The RP based on the 80% overlap is shown in Fig. S1(c). We observe that the RP avoids substantial tangential motion.

We repeated the same analysis using the other frequency bands (i.e., alpha, beta, and gamma), both the AEC and PLI as a functional connectivity measure, and the six network distance measures described in section 2.4. The results obtained for each combination were similar to those presented in Fig. S1. We also repeated this analysis for two other randomly chosen healthy controls and found similar results. Therefore, we decided to use the 80% overlap in the analyses presented in the main text.

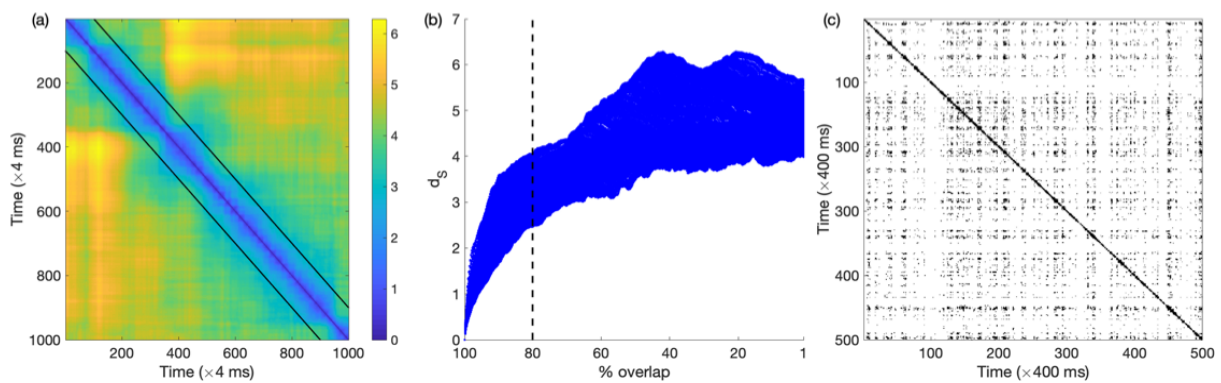

Figure S1: Assessment of the overlap between consecutive segments for the recurrence analysis. (a) Distance matrix based on the spectral norm ($d_S$), Eq. (5). A point ($a, b$) in this matrix is the spectral norm of the difference between networks inferred at the time windows $a$ and $b$. Points in between the two solid lines parallel to the main diagonal correspond to overlaps higher than 80%. (b) Spectral norm of the difference between pairs of networks as a function of the overlap. At large overlaps the distance is small, demonstrating the similarity between the networks. The dashed line at 80% overlap represents our choice. (c) Recurrence plot using the spectral norm and an overlap of 80%. Note that the temporal scale in panel (c) is different from that in panel (a); in panel (a) one time unit corresponds to 4 ms (i.e., overlap of $499/500$), whereas in panel (c) one time unit is 400 ms (i.e., overlap of $400/500$). We obtained these results using MEG data from one healthy individual filtered in the theta band. We employed the AEC to calculate the functional connectivity.

**Text S2: Functional network measures**

To construct functional networks from the MEG data, we used the phase lag index (PLI) (Stam et al., 2007) and the amplitude envelope correlation (AEC) with orthogonalized signals (Hipp et al., 2012).

Consider two sources corresponding to two nodes in the network. The PLI between the two nodes is given by

$$\text{PLI} = |\langle \text{sign}[\sin \Delta\phi(t_k)] \rangle|, \tag{S1}$$

where $\Delta\phi(t_k)$ is the instantaneous phase difference between the two time series at the two nodes and at time $t_k$, $k = 1, \ldots, K$ ($K$ is the total number of time points in the segment; we set $K = 500$), sign is the sign function, and $\langle \cdot \rangle$ is the average of the instantaneous phase difference over time.

To compute the AEC between pairs of source signals, we first orthogonalized the source signals to avoid spurious correlations due to source leakage (Hipp et al., 2012). The orthogonalization is given by

$$Y_{\perp X}(t, f) = \text{imag}\left(Y(t, f) \frac{X^*(t, f)}{|X(t, f)|}\right), \tag{S2}$$

where $X(t, f)$ and $Y(t, f)$ are two source signals in a frequency band $f$, $X^*(t, f)$ is the complex conjugate of $X(t, f)$, $Y_{\perp X}(t, f)$ is the orthogonalized $Y(t, f)$ with respect to $X(t, f)$, and $\text{imag}(x)$ stands for the imaginary part of $x$. The orthogonalization is also done in the opposite direction, from $Y(t, f)$ to $X(t, f)$, yielding $X_{\perp Y}(t, f)$. We then calculated the Pearson's correlation between $Y_{\perp X}(t, f)$ and $X(t, f)$, and between $Y(t, f)$ and $X_{\perp Y}(t, f)$, and took the average of the two values to make the AEC symmetric.

**Text S3: Orientation of the Fiedler vectors**

Algorithms to compute the normalized Fiedler vector $\vec{v}(t)$ do not differentiate between $\vec{v}(t)$ and $-\vec{v}(t)$. When comparing two Fiedler vectors to calculate the distance between two networks, their orientation is not arbitrary because opposite orientations lead to different distance values. To address this, we determined the orientation of the Fiedler vectors under the assumption that if two networks are similar, then their respective Fiedler vectors are likely to be similar.

First, we computed functional networks with the maximal overlap between consecutive segments of data, such that consecutive networks shared 499 samples out of the 500 samples. Due to the high overlap, the consecutive networks were expected to be similar to each other. Second, we computed the Fiedler vector $\vec{v}(1)$ of the functional network at time $t = 1$, and chose an arbitrary orientation. Third, we computed two possible Fiedler vectors with opposite orientations for the functional network at time $t = 2$. Fourth, we calculated the Euclidean distance between $\vec{v}(1)$ and the two possible Fiedler vectors at $t = 2$. Fifth, we selected the orientation of the Fiedler vector at time $t = 2$ such that the Fiedler vector with that orientation was the closer to $\vec{v}(1)$ in terms of the Euclidean distance. We refer to the vector with the selected orientation as $\vec{v}(2)$. Sixth, we computed the two possible Fiedler vectors at time $t = 3$ and decided its orientation by taking the vector that was the closer to $\vec{v}(2)$ in the Euclidean distance. We refer to the selected Fiedler vector as $\vec{v}(3)$. We repeated these steps to sequentially and uniquely determine the Fiedler vectors $\vec{v}(2), \vec{v}(3), ..., \vec{v}(T)$, where $T$ is the number of all functional networks. Note that our subsequent analysis only used a fraction of these $T$ Fiedler vectors. As described in the Supplementary Material text S1, we used an overlap of 80%, i.e. a shift of 100 samples out of 500 samples from one segment to the next. Therefore, we used the Fiedler vectors $\vec{v}(1), \vec{v}(101), \vec{v}(201)$, and so forth.

**Text S4: Distance measures**

As described in section 2.3, we considered six distance measures between pairs of networks to infer RPs. Three of these measures are described in the main text. In this section, we describe the other three measures: the log-Euclidean distance, the maximum norm between the Fielder vectors, and the cosine dissimilarity between the Fiedler vectors.

The log-Euclidean distance between networks is given by

$$d_{LE}(A(t_1), A(t_2)) = \|\log(A(t_1)) - \log(A(t_2))\|_F = \sqrt{\sum_{i=1}^{N}\sum_{j=1}^{N}(\log(a_{ij}(t_1)) - \log(a_{ij}(t_2)))^2}, \quad (S3)$$

with the convention of $\log(a_{ij}(t_a)) = 0$ when $a_{ij}(t_a) = 0$ (Arsigny et al., 2006). This distance measure was shown to be suitable for evaluating symmetric positive-definite matrices (Arsigny et al., 2006), which our functional networks are.

The maximum norm between the Fielder vectors is given by

$$d_{MF}(A(t_1), A(t_2)) = \max_{1 \le k \le N} |v_k(t_1) - v_k(t_2)|. \quad (S4)$$

The cosine dissimilarity between the Fiedler vectors is given by

$$d_{DF}(A(t_1), A(t_2)) = 1 - \vec{v}(t_1) \cdot \vec{v}(t_2). \quad (S5)$$

**Text S5: Reduction in the number of configurations**

To remove redundant configurations from our analysis, we studied the relations between the different configurations using MEG data from three arbitrarily selected healthy controls. First, for each configuration, participant, and threshold value, we computed the RP and then the 11 RQA values. Therefore, for each participant and threshold value, we obtained a matrix $X$ of size $(48 \times 11)$ that stored the 11 RQA values for each of the 48 configurations. Then, for a given participant and threshold value, we computed the Pearson's correlation coefficient between every pair of rows of the matrix $X$, i.e. the correlation between the two configurations by regarding the RQA values as samples. All pairs of configurations with the same functional connectivity measure and network-distance measure but different frequency bands had correlations larger than $0.9$. In other words, different frequency bands yielded similar information about recurrence. This result was consistent across the three participants and threshold values. Therefore, we decided to focus on only one frequency band. We chose the theta band because the original MEG recordings had the highest power in this band. By making this choice we reduced the space of configurations from 48 to 12 (i.e., 2 functional connectivity measures times 6 distance measures).

We then searched for a set of representative configurations such that all configurations had a correlation higher than $0.9$ to at least one of the configurations in the set for each of the three participants and the three thresholds. We tested all combinations of two, three and four configurations as potential candidates for a set of representative configurations. It turned out that such a set required at least four configurations. We found a set of four configurations with an average minimum correlation of $0.94$ to all other configurations across controls and thresholds comprising: (i) theta band + AEC + Frobenius norm ($d_F$), (ii) theta band + AEC + spectral norm ($d_S$), (iii) theta band + AEC + Euclidean distance between Fiedler vectors ($d_{EF}$), and (iv) theta band + PLI + spectral norm ($d_S$). Thus, each of the 48 initial configurations was represented by at least one of these four configurations in the sense that their pair correlation was at least $0.9$. We acknowledge that the choice of using a pair correlation of $0.9$ to decide whether two configurations yield equivalent recurrence information is arbitrary. Also, there is a chance that by neglecting 44 configurations based on this criterion we may be removing potentially informative configurations from our analysis. This, however, was a methodological choice with which we aimed to avoid negative consequences of keeping possibly redundant information in the subsequent analysis. For example, in section 3.1 of the main text we tested whether RQA measures were significantly different between the control group and the group with epilepsy across different configurations, while correcting for multiple testing across configurations. In this case, if different configurations are redundant, the use of all the 48 configurations would be detrimental to the power of each statistical test, while it would not provide additional information.

To test whether we neglected useful information by only taking the aforementioned four configurations, we ran an additional analysis. We considered an additional configuration, the theta+PLI+$d_{EF}$ configuration. We computed RPs and respective RQA measures for this fifth configuration for all individuals in the MEG data set (i.e., both controls and people with epilepsy). We used a threshold such that the density of points in the RPs was $0.05$. We found that the RQA measures obtained with this configuration had an average Pearson correlation of $0.976$ to the the RQA measures obtained with theta+AEC+$d_{EF}$ configuration, which is one of the four representative configurations used in the main text. Then, we tested the ability of the RQA measures to classify the two groups using the two configurations separately (see section 2.7 for details about the classification methods). The theta+AEC+$d_{EF}$ configuration yielded a classification accuracy of 69.2% (see Fig. 3(b)), and the theta+PLI+$d_{EF}$ configuration yielded 61.5%. We then combined the RQA measures from the two configurations, performed an additional classification, and obtained an accuracy of 65.4%. This result suggests that the theta+PLI+$d_{EF}$ configuration does not add significant, or at least relevant, information to theta+AEC+$d_{EF}$ configuration. We also expect that the other neglected configurations do not add sizable information to that obtained with the four representative configurations. This is a supposition whose validity should be tested in future work.

In the case of the sEEG data set, we initially had 6 configurations corresponding to the 6 distance measures. Following the results for the MEG data set, we focused on 3 configurations, i.e., those based on the Frobenius norm, the spectral norm, and the Euclidean distance between Fiedler vectors.

**Text S6: RQA measures**

We used 11 RQA measures to quantify RPs (Marwan et al., 2007; Marwan et al., 2015). For consistency with Marwan et al.'s notation, here we use $N \times N$ as the size of the RP, and the indices $i$ and $j$ to denote the entries of the recurrence matrix (i.e. $R_{i,j}$ instead of $R_{t_1,t_2}$ as in the main text).

Four measures were based on diagonal lines in the RP:

(1) The determinism ($DET$) is given by

$$DET = \frac{\sum_{l=l_{\min}}^{N} lP_1(l)}{\sum_{l=1}^{N} lP_1(l)}, \tag{S6}$$

where $l$ is the length of a diagonal line in the RP (i.e., a consecutive sequence of matrix entries equal to 1 that are parallel to the main diagonal in the recurrence matrix). The quantity $P_1(l)$ is the number of diagonal lines of length $l$ in the RP and is given by

$$P_1(l) = \sum_{i,j=1}^{N} (1 - R_{i-1,j-1})(1 - R_{i+l,j+l}) \prod_{k=0}^{l-1} R_{i+k,j+k}, \tag{S7}$$

where $R_{i,j}$ is the $(i,j)$th entry of the recurrence matrix, Eq. (2). The $DET$ is therefore the ratio of recurrence points that form diagonals in the RP of at least length $l_{\min}$ to all recurrence points. We used $l_{\min} = 2$. Note that a diagonal line of length $l$ means that pairs of dFNs at different times remained similar to each other for the duration of $l$ consecutive time points.

(2) The mean diagonal line length ($\langle L \rangle$) is defined as

$$\langle L \rangle = \frac{\sum_{l=l_{\min}}^{N} l P_1(l)}{\sum_{l=l_{\min}}^{N} P_1(l)}. \tag{S8}$$

(3) The maximal diagonal line length ($L_{\max}$) is the longest diagonal line in the RP, i.e.,

$$L_{\max} = \max\left(\{l_i\}_{i=1}^{N_l}\right), \tag{S9}$$

where $\{l_i\}$ is the set of all diagonal line lengths in the RP, and $N_l$ is the total number of diagonal lines, which is given by

$$N_l = \sum_{l \geq l_{\min}} P_1(l). \tag{S10}$$

(4) The entropy of the diagonal line length ($ENTR$) is given by

$$ENTR = -\sum_{l=l_{\min}}^{N} p(l) \ln p(l), \tag{S11}$$

where $p(l) = P_1(l)/N_l$, i.e. the probability to find a diagonal line of length $l$. Thus, $ENTR$ is the Shannon entropy of $p(l)$ and quantifies the complexity of the RP in terms of diagonal lines.

Additionally, we used the following three measures based on vertical lines in the RP:

(1) The laminarity ($LAM$) is analogous to the determinism and is defined as the ratio of recurrence points that form vertical lines to all recurrence points, i.e.

$$LAM = \frac{\sum_{v=v_{\min}}^{N} v P_2(v)}{\sum_{v=1}^{N} v P_2(v)}, \tag{S12}$$

where $v$ is the length of vertical lines in the RP, $P_2(v)$ is the number of vertical lines of length $v$ in the RP, which is given by

$$P_2(v) = \sum_{i,j=1}^{N} (1 - R_{i,j-1})(1 - R_{i,j+v}) \prod_{k=0}^{v-1} R_{i,j+k}, \tag{S13}$$

and $v_{min}$ is the minimum length of vertical lines to be considered. We used $v_{min} = 2$.

(2) The trapping time ($TT$) is the average length of vertical lines and is given by

$$TT = \frac{\sum_{v=v_{min}}^{N} v P_2(v)}{\sum_{v=v_{min}}^{N} P_2(v)}. \tag{S14}$$

(3) The maximal vertical line length ($V_{max}$) is defined as

$$V_{max} = \max\big(\{v_i\}_{i=1}^{N_v}\big), \tag{S15}$$

where $\{v_i\}$ is the set of all vertical line lengths in the RP, and $N_v$ is the total number of vertical lines given by

$$N_v = \sum_{v \geq v_{min}} P_2(v). \tag{S16}$$

Furthermore, we considered three measures that assess recurrence times. For a given column $j$ in the recurrence matrix, the elements $R_{i,j} = 1, 1 \leq i \leq N$, are its recurrence points. A recurrence time of first type of column $j$ is the number of time points from one recurrence point to the next along the column. Consecutive recurrence points, i.e. $R_{i,j} = 1$ and $R_{i,j+1} = 1$, are counted as a recurrence time of first type equal to 1. To calculate the overall recurrence time of first type of the RP, denoted by $T1$, we identify all recurrence times of first type across all columns and then average them.

Recurrence times equal to 1 may result from tangential motion and not actual recurrences of the dynamical system. To account for this, one can alternatively define recurrence times as the length of time between recurrence points neglecting recurrence times of 1. The recurrence time of second type, denoted by $T2$, is the average of all such recurrence times in the RP.

The recurrence time entropy ($RTE$) is given by

$$RTE = -\frac{1}{\ln V'_{max}} \sum_{v=1}^{V'_{max}} H(v) \ln H(v), \tag{S17}$$

where $V'_{max}$ is the maximum length of vertical empty lines in the RP, and $H(v)$ is the distribution of the length of vertical empty lines. A vertical empty line is a consecutive sequence of matrix entries equal to 0 along a column in the recurrence matrix.

Finally, we considered the transitivity ($Trans$) of the RP, which is defined as

$$Trans = \frac{\sum_{i,j,k=1}^{N} R_{j,k} R_{i,j} R_{i,k}}{\sum_{i,j,k=1}^{N} R_{i,j} R_{i,k}(1 - \delta_{j,k})}. \tag{S18}$$

**Text S7: Classification of people with epilepsy and healthy controls using the weighted mean degree**

We asked whether the weighted mean degree of the dFNs, which is a simpler measure of functional connectivity than those based on RQA measures, is capable of classifying people with epilepsy and controls. As we did in the RQA-based classification shown in the main text, we considered the dFNs obtained in the theta band. For each individual and each functional connectivity measure (i.e., AEC or PLI), we calculated the weighted mean degree of each of the 506 functional networks. The weighted mean degree $\langle w(t) \rangle$ at time $t$ was given by (Rubinov and Sporns, 2010):

$$\langle w(t) \rangle = \frac{2}{90 \times 89} \sum_{i=1}^{90} \sum_{j=1}^{i-1} a_{ij}(t). \tag{S19}$$

Then, we computed the time average of the weighted mean degree, i.e.,

$$W = \frac{1}{506} \sum_{t=1}^{506} \langle w(t) \rangle. \tag{S20}$$

We then compared the $W$ values between the two groups of individuals and observed that they were not statistically different for each functional connectivity measure (i.e., AEC or PLI) (Mann-Whitney $U$ test). Finally, using the two $W$ values as features to classify the two groups, we found an accuracy of 67.3%; we used the cosine kNN classifier which was the best classifier as identified by MATLAB's Classification Learner Toolbox.

**Text S8: Classification of people with epilepsy and healthy controls using the RQA on RPs directly applied to multivariate MEG time series**

We compared our framework with the traditional RPs and RQA, i.e., those applied to MEG time series from individual sources instead of to dFNs. Because we only considered the theta band in the main text, we also restricted the traditional approach to the time series in the theta band. For each individual, we computed a traditional RP using Eq. (1) with

$$d(\vec{x}(t_a), \vec{x}(t_b)) = \sqrt{\sum_{k=1}^{90} (x_k(t_a) - x_k(t_b))^2}, \tag{S21}$$

where $\vec{x}(t_a) = \left(x_1(t_a), x_2(t_a), \ldots, x_{90}(t_a)\right)^\top$ and $\vec{x}(t_b) = \left(x_1(t_b), x_2(t_b), \ldots, x_{90}(t_b)\right)^\top$ were vectors from the source reconstructed signals at times $t_a$ and $t_b$, respectively. Note that, whereas in the dFNs framework we used 500 samples to generate a network and an overlap of 80% between consecutive networks, here we used every sample as an independent time point. Therefore, the traditional RP was much larger than an RP for dFNs. For computational tractability, we restricted the size of the RP to $5000 \times 5000$ constructed from the first 5000 samples. We set the threshold on the distance values to define recurrence to 0.05, which was the same value as the one used in the main text. Next, we used the RQA measures to quantify the RPs of each individual. Finally, we employed the RQA measures to classify people as healthy or having epilepsy and obtained an accuracy of 69.2%; we used the linear support vector machine which was the best classifier as identified by MATLAB's Classification Learner Toolbox.

**Text S9: A note on the RQA measures that differ between people with epilepsy and healthy controls in the MEG data set**

We found that both the recurrence times of first ($T1$) and second type ($T2$) are smaller in people with epilepsy compared to healthy controls. However, this finding was specific to the AEC+$d_F$ and AEC+$d_S$ configurations. In the AEC+$d_{EF}$ configuration we found a significantly higher $T1$ in people with epilepsy but no statistical difference in $T2$ between the two groups. Given that the difference between $T1$ and $T2$ is that $T1$ is affected by tangential motion, this finding implies that RPs from controls in the AEC+$d_{EF}$ configuration reflect more tangential motion than those from people with epilepsy. We also found that the trapping time ($TT$) was significantly smaller in people with epilepsy than controls in the PLI+$d_S$ configuration (see Fig. 2(f)). The $TT$ is the average length of vertical lines in the RP and measures the time during which dFNs are trapped near a certain network. In this way, PLI networks from healthy controls are more likely to be trapped than those from people with epilepsy. In contrast, the $TT$ of AEC networks is not significantly different between the two groups in any of the three considered configurations. The fact that we observed different results for different configurations is not a contradiction, because different configurations assess different features of the functional networks.

**Additional Supporting Figures and Table**

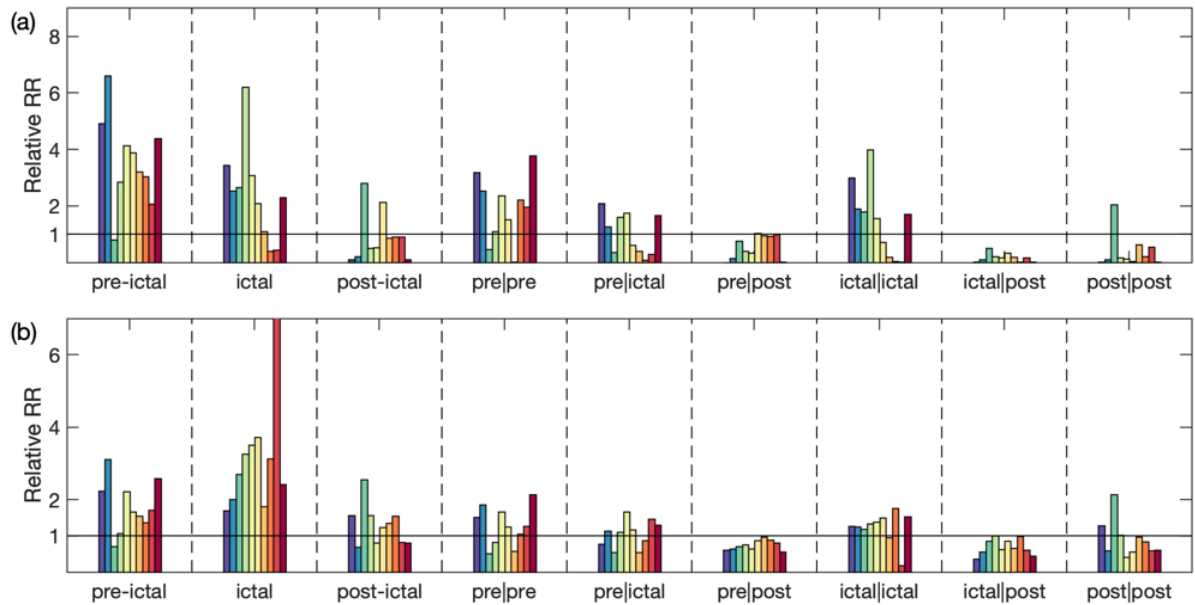

Figure S2: Relative recurrence rate ($RR$) for the different blocks of the RP from sEEG data. Panel (a) and (b) are equivalent to Fig. 4(c), except that to compute the relative $RR$, we constructed the RPs using (a) the spectral norm, Eq. (4), and (b) the Euclidean norm between Fiedler vectors, Eq. (5). The 9 different types of blocks in the RP are described in the caption of Fig. 4(c). As in Fig. 4(c), for each type of block, we plot 10 bars of different colors, one for each individual. To obtain the RPs, we used a density of recurrence points of $0.05$.

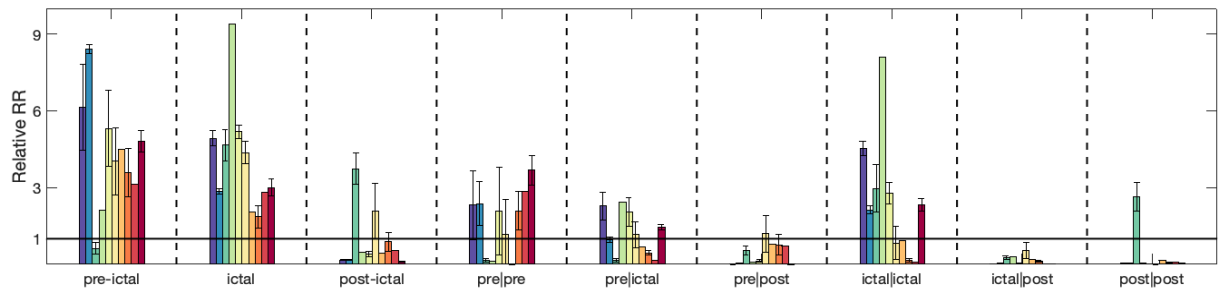

Figure S3: Relative recurrence rate ($RR$) for the different blocks of the RP from sEEG data. This figure is equivalent to Fig. 4(c), except that we only used two peri-ictal epochs per individual to compute the relative $RR$. For individuals with more than two peri-ictal epochs, we computed the relative $RR$ for all possible combinations of two peri-ictal epochs and plotted the average by the bars. The error bars represent the standard error. The bars that do not have an error bar, such as the fourth bar from the left on each block, correspond to the individuals for which we had only two peri-ictal epochs.

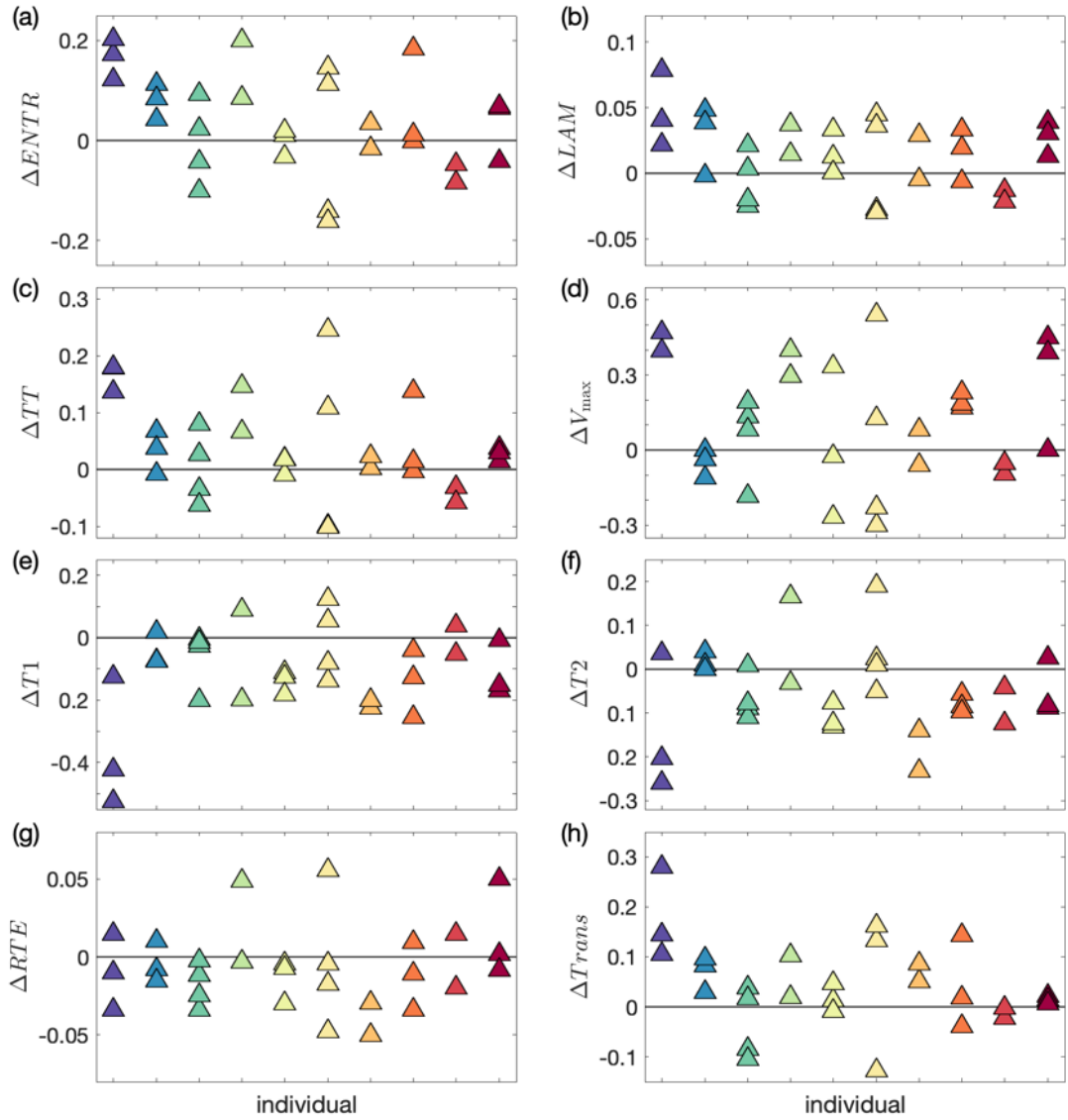

Figure S4: Variation in RQA values between pre- and post-ictal RPs of sEEG dFNs. Each panel represents the variation in one RQA measure, as in Figs. 7(c)-(e). In each panel, each triangle corresponds to a different peri-ictal epoch, and the triangles of the same color belong to the same individual. All the observed tendencies significantly deviated from zero (Wilcoxon signed-rank test at a significance level of $0.05$). To compute the underlying RPs, we used the Frobenius norm as a pairwise distance between networks, Eq. (3), and a density of recurrence points of $0.05$ to define the threshold $\epsilon$.

|                       | AEC+$d_S$(T2) | AEC+$d_F$(all) | AEC+$d_S$(all) | AEC+$d_{EF}$(all) | PLI+$d_S$(all) | PCA  |
|-----------------------|---------------|----------------|----------------|-------------------|----------------|------|
| **Fine Tree**         | 61.5          | 61.5           | 59.6           | 48.1              | 50.0           | 59.6 |
| **Medium Tree**       | 61.5          | 61.5           | 59.6           | 48.1              | 50.0           | 59.6 |
| **Coarse Tree**       | 63.5          | **65.4**       | 61.5           | 57.7              | 34.6           | **63.5** |
| **Linear Discriminant** | **69.2**    | 46.2           | 59.6           | 50.0              | 59.6           | 61.5 |
| **Quadratic Discriminant** | 65.4    | 50.0           | 44.2           | 50.0              | 63.5           | 36.5 |
| **Logistic Regression** | **69.2**    | 46.2           | 55.8           | 48.1              | 57.7           | 61.5 |
| **Gaussian Naïve Bayes** | 65.4       | 55.8           | 61.5           | 67.3              | 61.5           | 61.5 |
| **Kernel Naïve Bayes** | 63.5         | 51.9           | 65.4           | 63.5              | 57.7           | 50.0 |
| **Linear SVM**        | 63.5          | 61.5           | 63.5           | 55.8              | 63.5           | 51.9 |
| **Quadratic SVM**     | 65.4          | 55.8           | 55.8           | 48.1              | 53.8           | 50.0 |
| **Cubic SVM**         | 40.4          | 59.6           | 57.7           | 51.9              | 57.7           | 48.1 |
| **Fine Gaussian SVM** | 46.2          | 17.3           | 25.0           | 53.8              | 17.3           | 3.8  |
| **Medium Gaussian SVM** | 67.3        | 63.5           | 67.3           | 61.5              | 55.8           | 46.2 |
| **Coarse Gaussian SVM** | 67.3        | 32.7           | 51.9           | 61.5              | 57.7           | 5.8  |
| **Fine KNN**          | 65.4          | 55.8           | 51.9           | 51.9              | **65.4**       | 61.5 |
| **Medium KNN**        | 63.5          | 51.9           | **71.2**       | 65.4              | 55.8           | 50.0 |
| **Coarse KNN**        | 3.8           | 3.8            | 3.8            | 3.8               | 3.8            | 3.8  |
| **Cosine KNN**        | 50.0          | 51.9           | 67.3           | **69.2**          | 61.5           | 44.2 |
| **Cubic KNN**         | 63.5          | 51.9           | 69.2           | 61.5              | 51.9           | 38.5 |
| **Weighted KNN**      | 65.4          | 59.6           | 67.3           | 61.5              | 61.5           | 51.9 |
| **Boosted Trees**     | 3.8           | 3.8            | 3.8            | 3.8               | 3.8            | 3.8  |
| **Bagged Trees**      | 65.4          | 53.8           | 59.6           | 61.5              | 40.4           | 53.8 |
| **Subspace Discriminant** | **69.2**  | 59.6           | 63.5           | 57.7              | 61.5           | **63.5** |
| **Subspace KNN**      | 65.4          | 57.7           | 55.8           | 48.1              | 59.6           | 57.7 |
| **RUSBoosted Trees**  | 34.6          | 28.8           | 32.7           | 34.6              | 21.2           | 21.2 |

Table S1: Accuracy of the classification of people with epilepsy using RQA applied to dFNs inferred from MEG data. Each row corresponds to a classifier from MATLAB's Classification Learner Toolbox. Each column corresponds to a set of features used for the classification: the recurrence time of second type (T2) from the AEC+$d_S$ configuration, all RQA measures from each of the four representative configurations, and 12 principal components explaining 85% of the variance of all RQA measures across the four configurations. We applied a 50-fold cross-validation procedure to avoid over-fitting. SVM stands for support vector machine.